\documentclass[11pt]{article}

\usepackage[indentafter]{titlesec}
\titleformat{name=\section}{}{\thetitle.}{0.8em}{\centering\scshape}
\titleformat{name=\subsection}{}{\thetitle.}{0.5em}{\centering\itshape}
\titleformat{name=\subsubsection}[runin]{}{\thetitle.}{0.5em}{\itshape}[.]
\titleformat{name=\paragraph,numberless}[runin]{}{}{0em}{}[.]
\titlespacing{\paragraph}{0em}{0em}{0.5em}
\titleformat{name=\subparagraph,numberless}[runin]{}{}{0em}{}[.]
\titlespacing{\subparagraph}{0em}{0em}{0.5em}

\usepackage{caption}
\usepackage{amssymb,amsmath,amsfonts}
\usepackage{enumerate}
\usepackage[ruled]{algorithm2e}

\usepackage{graphicx}        

\usepackage[bottom]{footmisc}
\usepackage{url}

\usepackage{xargs}

\usepackage[]{xcolor}
\definecolor{dblue}{rgb}{0,.1,.5}  

\usepackage[round,semicolon]{natbib}

\usepackage[]{hyperref}

\usepackage[amsmath,thmmarks]{ntheorem}
\theoremstyle{break}
\renewtheorem{procedure}{Procedure}

\usepackage[figure,table]{hypcap} 
\hypersetup{
    pdftoolbar=true,        
    pdfmenubar=true,        
    pdffitwindow=true,     
    pdfstartview={FitH},    
    pdftitle={},           
    pdfauthor={},         
    pdfsubject={},      
    pdfcreator={},      
    colorlinks=true,        
    linkcolor=dblue,          
    citecolor=dblue,        
     urlcolor=dblue,      
}

\makeatletter
\def\url@leostyle{%
  \@ifundefined{selectfont}{\def\UrlFont{\sf}}{\def\UrlFont{\color{dblue}\small\sffamily}}}
\makeatother
\begin{document}

\title{An Approach to Efficient Fitting of Univariate and Multivariate Stochastic Volatility Models}
\author{Chen Gong$^*$ \\
Department of Statistics, University of Pittsburgh \\ 
\small\url{chg87@pitt.edu} \and 
 David S. Stoffer\,\thanks{This work was supported in part by NSF DMS-1506882} \\
 Department of Statistics, University of Pittsburgh\\
\small\url{stoffer@pitt.edu} }

\date{}
\maketitle


\begin{abstract} 
The  stochastic volatility   model  is a popular tool for modeling the volatility of assets. 
The model is  a nonlinear and non-Gaussian state space model, and consequently is difficult to fit. 
Many approaches, both classical and Bayesian, have been developed that rely on numerically
intensive techniques such as quasi-maximum likelihood estimation and Markov chain Monte Carlo (MCMC).
Convergence and mixing  problems still plague MCMC algorithms when drawing samples
sequentially from the posterior distributions.
While particle Gibbs methods have been successful  when applied to nonlinear or non-Gaussian
state space models in general, slow convergence still haunts the technique when applied specifically
to stochastic volatility models. 
We present an approach  that couples particle Gibbs with ancestral sampling
and joint parameter sampling
that ameliorates the slow convergence and mixing problems when fitting both univariate and multivariate
 stochastic volatility models.  We demonstrate the enhanced method on various numerical examples.
\end{abstract}

\maketitle


\def\Cref{\autoref}
\def\bm{\pmb}

\newcommand{\chunk}[3]{#1_{#2:#3}}
\newcommand{\E}{\ensuremath{\operatorname{e}}}
\def\half{{\tfrac{1}{2}}}
\def\pr{\mathbb P}
\def\P{{\rm P}}
\def\corr{\PCor}               
\def\|{\,|\,}
\def\tint{{\ensuremath{\textstyle\int}}}
\def\eb{\epsilon}
\def\e{{\epsilon}}
\def\s{{\sigma}}
\def\bb{{\mb\beta}}
\def\ab{{\mb\alpha}}
\def\pb{{\mb\phi}}
\def\l{\lambda}
\def\rb{{\mb\rho}}
\newcommand{\N}{\ensuremath{\operatorname{N}}}
\newcommand{\p}{\ensuremath{\operatorname{p}}}
\newcommand{\q}{\ensuremath{\operatorname{q}}}
\def\T{n}
\newcommand{\dens}{\ensuremath{\operatorname{p}}}

\def\m{{\mb m}}
\def\r{{\mb r}}
\def\h{{\mb h}}
\def\y{{Y}}
\def\w{{W}}
\def\a{{\mb a}}
\def\B{{\cal B}}
\def\v{{V}}
\def\u{{U}}

\def\b{{\mb b}}
\def\th{{\theta}}

\def\mub{{\mu_0}}
\def\tr{{\rm tr\,}}

\def\bpm{\begin{pmatrix}}
\def\epm{\end{pmatrix}}
\def\bma{\begin{matrix}}
\def\ema{\end{matrix}}

\def\R{{\sf R}}


\newcommand{\argmax}{\operatornamewithlimits{arg\,max}}
\newcommand{\argmin}{\operatornamewithlimits{arg\,min}}

\def\rset{\ensuremath{\mathbb{R}}}
\def\qset{\ensuremath{\mathbb{Q}}}
\def\nset{\ensuremath{\mathbb{N}}}
\def\zset{\ensuremath{\mathbb{Z}}}
\def\cset{\ensuremath{\mathbb{C}}}


\newcommand{\rf}[1]{(\ref{#1})}    
\newcommand{\rff}[1]{(\ref*{#1})}    
\newcommand{\Z}{\ensuremath{Z}}     
\newcommand{\z}{\ensuremath{z}}
\newcommand{\be}{\begin{equation}}     
\newcommand{\ee}{\end{equation}}        
\def\t{{\pmb t}}
\newcommand{\distiid}{\ensuremath{\stackrel{\mathrm{iid}}{\sim\,}}}    
\newcommand{\simiid}{\ensuremath{ {\sim}\,{\rm iid\,} }}         
\newcommand{\wn}{\ensuremath{\mathrm{WN}}}   
\newcommand{\om}{\omega}
\newcommand{\omj}{\omega_j}
\newcommand{\predxmse}[2]{P_{#1|#2}}  
\newcommand{\mslim}{\ensuremath{\stackrel{\text{m.s.}}{\longrightarrow}}}   
\newcommand{\PCor}{\ensuremath{\operatorname{Cor}}}  
\newcommand{\cum}{{\rm cum}}
\newcommand{\doublesum}{\mathop{\sum\sum}}
\newcommand{\triplesum}{\mathop{\sum\sum\sum}}  
\newcommand{\dotsum}{\mathop{\sum\dots\sum}}

\newcommand{\Tr}[1]{\mathrm{Tr}(#1)}
\newcommand{\proj}{\ensuremath{{\sf P\!}}}
\newcommand{\mcy}{\ensuremath{\mathcal Y}}
\newcommand\pdi{\ \partial_i}
\newcommand\pdj{\ \partial_j}


\providecommand{\abs}[1]{\lvert#1\rvert}
\newcommand{\Xinit}{\xi} 
\def\borel{\mathcal{B}}
\def\bepsilon{\boldsymbol{\epsilon}}
\def\lleb{\ensuremath{\mathrm{Leb}}}
\def\bx{\bm{x}}
\def\by{\bm{y}}
\def\bz{\bm{z}}
\def\bX{\bm{X}}
\def\bY{\bm{Y}}
\def\bZ{\bm{Z}}
\newcommand{\vvec}[1]{\bm{#1}}
\newcommand{\VVec}[1]{\mathrm{Vec}(#1)}
\def\bM{\mathbf{M}}
\newcommand{\coint}[1]{\left[#1\right)}
\newcommand{\ocint}[1]{\left(#1\right]}
\newcommand{\ooint}[1]{\left(#1\right)}
\newcommand{\ccint}[1]{\left[#1\right]}
\newcommand{\Id}{\ensuremath{\mathrm{I}}}

\newcommand{\closure}[1]{\bar{#1}}
\newcommand{\cas}{{\:\stackrel{\prob-\mathrm{a.s.}}{\longrightarrow}\:}}
\newcommand{\cp}{{\:\stackrel{\prob}{\longrightarrow}\:}}

\def\epartsymb{x}
\newcommand{\epart}[2][]
{%
\ifthenelse{\equal{#1}{}}{\ensuremath{\epartsymb^{#2}}}{\ensuremath{\epartsymb^{#2}_{#1}}}
}
\newcommand{\mpart}[3]
{
\epartsymb^{#3}_{#1:#2}
}

\newcommand{\etpart}[2][]
{%
\ifthenelse{\equal{#1}{}}{\ensuremath{\tilde{\epartsymb}^{#2}}}{\ensuremath{\tilde{\epartsymb}^{#2}_{#1}}}
}
\newcommand{\etwght}[2][]{%
\ifthenelse{\equal{#1}{}}{\ensuremath{\tilde{\omega}^{#2}}}{\ensuremath{\tilde{\omega}^{#2}_{#1}}}
}
\newcommand{\ewght}[2][]{%
\ifthenelse{\equal{#1}{}}{\ensuremath{\omega^{{#2}}}}{\ensuremath{\omega^{{#2}}_{{#1}}}}
}
\newcommand{\wwght}[2][]{%
\ifthenelse{\equal{#1}{}}{\ensuremath{w^{{#2}}}}{\ensuremath{w^{{#2}}_{{#1}}}}
}
\newcommand{\NISE}[4][]%
{%
\ifthenelse{\equal{#1}{}}{\ensuremath{\tilde{#2}^{\scriptstyle \mathrm{IS}}_{#4}\left(#3 \right)}}{\ensuremath{\tilde{#2}^{\scriptstyle \mathrm{IS}}_{#1,#4}\left(#3 \right)}}%
}
\newcommand{\ISE}[4][]%
{%
\ifthenelse{\equal{#1}{}}{\ensuremath{\widehat{#2}^{\scriptstyle  \mathrm{IS}}_{#4} \left(#3 \right)}}{\ensuremath{\widehat{#2}^{\scriptstyle  \mathrm{IS}}_{#1,#4} \left(#3 \right)}}%
}
\newcommand{\SIRE}[4][]%
{%
\ifthenelse{\equal{#1}{}}{\ensuremath{\hat{#2}^{\scriptstyle  \mathrm{SIR}}_{#4} \left(#3 \right)}}{\ensuremath{\hat{#2}^{\scriptstyle  \mathrm{SIR}}_{#1,#4} \left(#3 \right)}}%
}
\newcommandx\Kun[2][1=]{\ensuremath{Q^{#1}}_{#2}}
\newcommandx\bKun[2][1=]{\ensuremath{\bar{Q}^{#1}}_{#2}}
\newcommand{\kun}[1]{\ensuremath{q}_{#1}}

\newcommandx{\kiss}[2][1=]
{\ifthenelse{\equal{#1}{}}{r_{#2}}
{\ifthenelse{\equal{#1}{fully}}{p^{\star}_{#2}}
{\ifthenelse{\equal{#1}{smooth}}{\tilde{r}_{#2}}{\mathrm{erreur}}}}}

\newcommandx{\Kiss}[3][1=,3=]
{\ifthenelse{\equal{#1}{}}{R_{#2}^{#3}}
{\ifthenelse{\equal{#1}{fully}}{P^{\star}_{#2}}
{\ifthenelse{\equal{#1}{smooth}}{\tilde{R}^{#3}_{#2}}{\mathrm{erreur}}}}}

\newcommand{\XinitTAR}{\ensuremath{\phi_{0}}}
\newcommand{\TAR}[1]{\ensuremath{\phi_{0:#1|#1}}}
\newcommand{\logl}[2][]%
{%
\ifthenelse{\equal{#1}{}}{\ensuremath{\ell_{#2}}}{\ensuremath{\ell_{#1,#2}}}%
}

\def\KJointHMM{K}
\def\kJointHMM{k}

\newcommand{\XinitIS}{\ensuremath{r_0}}
\newcommand{\ETAR}[1]{\ensuremath{\hat{\phi}_{0:#1|#1}}}
\newcommandx\weightfunc[2][1=,2=]
{\ifthenelse{\equal{#2}{}}{\omega_{#1}}{\omega^{#2}_{#1}}}

\newcommand{\sigmaopt}[1]{\tilde{\sigma}_{#1}}
\newcommand{\meanopt}[1]{\tilde{m}_{#1}}
\newcommand{\NormOPT}{\ensuremath{\gamma}}
\newcommand{\WRoot}{\ensuremath{R}}
\newcommand{\VRoot}{\ensuremath{S}}
\newcommand{\VCov}[1][]%
{%
\ifthenelse{\equal{#1}{}}{\VRoot \VRoot^t}{\VRoot_{#1} \VRoot^t_{#1}}%
}
\newcommand{\KGF}[1]{\ensuremath{K_{#1}}}
\def\CV{\mathrm{CV}}
\newcommand{\ebwght}[2]{\ensuremath{\bar{\omega}_{#1}^{#2}}}
\newcommand{\proppart}[2]{\ensuremath{\tilde{\epartsymb}_{#1}^{#2}}}
\newcommand{\PROPTAR}[1]{\ensuremath{\tilde{\phi}_{0:#1|#1}}}
\newcommand{\sumweight}[2][]{%
\ifthenelse{\equal{#1}{}}{\ensuremath{\Omega^{#2}}}{\ensuremath{\Omega_{#1}^{#2}}}}
\newcommand{\epartset}{\ensuremath{\mathbb{E}}}
\newcommand{\epartsigma}{\ensuremath{\mathcal{E}}}
\newcommand{\etpartset}{\ensuremath{\mathbb{\tilde{E}}}}
\newcommand{\etpartsigma}{\ensuremath{\mathcal{\tilde{E}}}}
\def\stdnormfunc{\sigma}
\newcommand\ntimes[1]{M^{#1}}
\newcommand\ntimesres[1]{H_{#1}}
\newcommand{\partfrac}[1]{\left\langle #1 \right\rangle}
\newcommandx{\mcfpart}[2][1=]{%
\ifthenelse{\equal{#1}{}}{{\mathcal{F}^{#2}}}{\mathcal{F}_{#1}^{#2}}%
}
\newcommand{\Afunc}{\ensuremath{a}}
\newcommand{\Bfunc}{\ensuremath{b}}
\newcommand{\PX}{\ensuremath{X}}
\newcommand{\px}{\ensuremath{w}}
\newcommand{\PXset}{\ensuremath{\mathbb{W}}}
\newcommand{\PXsigma}{\ensuremath{\mathcal{W}}}
\newcommand{\DX}{\ensuremath{I}}
\newcommand{\dx}{\ensuremath{i}}
\newcommand{\DXset}{\ensuremath{\mathbb{I}}}
\newcommand{\DXsigma}{\ensuremath{\mathcal{I}}}
\newcommand{\order}{\ensuremath{r}}
\newcommand{\tsumweight}[2][]{%
\ifthenelse{\equal{#1}{}}{\ensuremath{\widetilde{\Omega}^{#2}}}{\ensuremath{\widetilde{\Omega}_{#1}^{#2}}}}

\newcommandx{\BK}[2][2=]{B^{#2}_{#1}}
\newcommandx{\bk}[2][2=]{b^{#2}_{#1}}

\newcommand{\esssup}[2][]%
{\ifthenelse{\equal{#1}{}}{| #2 |_\infty}{| #2 |^{#1}_{\infty}}}
\newcommandx{\F}[2][2=]{F_{#1}^{#2}}
\newcommand{\cfunc}{\bm{1}}

\newcommand{\asymVar}[4][]{
\ifthenelse{\equal{#1}{}}{\ifthenelse{\equal{#4}{}}{\ensuremath{\Gamma_{#2|#3}}}{\ensuremath{\Gamma_{#2|#3}\left[#4\right]}}}
{\ifthenelse{\equal{#4}{}}{\ensuremath{\Gamma_{#1,#2|#3}}}{\ensuremath{\Gamma_{#1,#2|#3}\left[#4\right]}}}
}
\newcommand{\incrasymVar}[4][]{
\ifthenelse{\equal{#1}{}}{\ensuremath{\sigma^2_{#2,#3}\left[#4\right]}}{\ensuremath{\sigma^2_{#1,#2,#3}\left[#4\right]}}
}

\def\param{\Theta, \mu}
\def\vparam{\vartheta}
\def\dimparam{{d}}
\newcommand{\HEM}[2]{\ensuremath{\mathcal{H}(#1 \, ; #2)}}
\newcommand{\oparam}{\ensuremath{\theta'}}
\newcommand{\ooparam}{\ensuremath{\theta''}}
                                       \newcommand{\Param}{\ensuremath{\Theta}}
\newcommand{\interior}[1]{{#1}^o}

\newcommand{\fisher}{\ensuremath{\mathcal{J}}}
\newcommand{\contrast}{\ensuremath{\ell}}
\newcommand{\mleparam}{\ensuremath{\widehat{\param}}}
\newcommand{\spsemi}{\, ;}
\newcommand{\statlogl}[1]{\ensuremath{\ell_{#1}^s}}
\newcommand{\plogl}{\ensuremath{\log \operatorname{L}}}

\newcommand{\dparam}{\ensuremath{d}}
\newcommand{\tparam}{\ensuremath{{\theta_\star}}}
\def\paramset{\Theta}
\renewcommandx{\m}[1][1=]
{\ifthenelse{\equal{#1}{}}{\ensuremath{m}}{\ensuremath{m^{#1}}}}
\newcommandx{\M}[1][1=]
{\ifthenelse{\equal{#1}{}}{\ensuremath{M}}{\ensuremath{M^{#1}}}}

\newcommandx{\g}[1][1=]
{\ifthenelse{\equal{#1}{}}{\ensuremath{g}}{\ensuremath{g^{#1}}}}

\newcommand{\plhood}{\operatorname{L}}

\newcommandx{\lhood}[3][1=]%
{\operatorname{L}_{#1}({#2};\, {#3})
}

\newcommandx{\hlhood}[4][1=,4=]%
{\widehat{\operatorname{L}}^{#4}_{#1}({#2};\, {#3})
}


\newcommandx{\hdens}[3][1=,2=]%
{
\ifthenelse{\equal{#1}{}}
{\ifthenelse{\equal{#2}{}}{\hat{p}(#3)}{\hat{p}^{#2}(#3)}}
{\ifthenelse{\equal{#2}{}}{\hat{p}_{#1}(#3)}{\hat{p}_{#1}^{#2}(#3)}}
}

\newcommandx{\QEM}[3][3=]{\ensuremath{\mathcal{Q}_{#3}(#1 \, ; #2)}}
\newcommandx{\hQEM}[3][3=]{\ensuremath {\hat{\mathcal{Q}}_{#3}(#1 \| #2)}}
\newcommand{\ls}{L}


\newcommandx{\deriv}[1][1=]{\nabla_{#1}}

\newcommandx\lkdM[3][1=,3=]{
\ifthenelse{\equal{#2}{}}
{ \mathsf{L}_{#1}^{#3}}
{ \mathsf{L}_{#1}^{#3}(#2)}
}
\newcommandx\lkdMStat[3][1=,3=]{
\ifthenelse{\equal{#2}{}}
{ \bar{\mathsf{L}}_{#1}^{#3}}
{ \bar{\mathsf{L}}_{#1}^{#3}(#2)}
}

\newcommandx\lkd[3][1=,3=]{
\ifthenelse{\equal{#2}{}}
{ \ell_{#1}^{#3}}
{ \ell_{#1}^{#3}\langle #2\rangle}
}
\newcommandx\lkdStat[3][1=,3=]{
\ifthenelse{\equal{#2}{}}
{ \bar \ell_{#1}^{#3}}
{ \bar \ell_{#1}^{#3}(#2)}
}

\newcommand{\mlStat}[1]{\bar \theta_{#1}}
\newcommand{\ml}[1]{\hat{\theta}_{#1}}
\newcommand{\thv}{{\theta_\star}}

\renewcommand{\-}{\mbox{-}}
\newenvironment{subproof}[1]
{\textbf{\em #1}.}{\hfill $\blacktriangleleft$ \newline \indent}

\newcommand{\pscal}[2]{\{ #1 \}' #2}
\newcommand{\hint}{\textit{Hint: }}

\newcommand{\CPEu}[3][]
{\ifthenelse{\equal{#1}{}}{{\mathbb E}\left[\left. #2 \, \right| #3 \right]}{{\mathbb E}^{#1}\left[\left. #2 \, \right | #3 \right]}}

\newcommandx{\f}[2][1=\theta]{f^{#1}_{#2}}
\newcommand{\mlY}[1]{\theta_{#1}}

\def\bigone{\mathbf{1}}

\newcommand\pgaskernel[1]{K^{#1}}
\def\Xcard{r}

%
\newcommand\Law{\mathcal{L}^N}
\newcommand\Mp{M}                                
\newcommand{\range}[2]{#1, \, \dots, \, #2}      
\newcommand{\crange}[2]{\{#1, \, \dots, \, #2\}} 
\newcommand{\prange}[2]{(#1, \, \dots, \, #2)}   
\newcommand\mcmcindex{k}
\newcommand\mcmcfinalindex{K}
\newcommand\lhest[2]{\widehat{\mathrm{L}}_{#1}^{#2}}
\newcommand\XX{\mathbf{X}}
\newcommand\II{\mathbf{I}}
\newcommand\xx{\mathbf{x}}
\newcommand\ii{\mathbf{i}}
\newcommand\postdens[2]{\phi_{#1\mid#2}}
\newcommand\jointprop[1]{p_{#1}}
\newcommand\apfpsi{\psi_n^N}
\newcommand\exttarget{\pi_n^N}
\newcommand\extendedspace{\Xi}


\section{The Problem}\label{sec:problem}

Most models of volatility that are used in practice are of a multiplicative
form, modeling the return of an asset, say $y_t$,  
observed at discrete time points, $t=1, \dots, \T$, as 
\be\label{eq:volatility}
y_t = \sigma_t \epsilon_t
\ee
where $\epsilon_t$ is an iid sequence  and the volatility process
$\sigma_t$   is a
non-negative stochastic process such that $\epsilon_t$ is independent of $\s_s$ for all $s \le t$.
It is often assumed that $\e_t$ has zero mean and unit variance.

The basic univariate discrete-time stochastic volatility (SV) model
 writes the returns and the stationary log volatility, $x_t = \log \s_t^2$, as
\begin{align}
x_t &= \mu + \phi(x_{t-1}-\mu) + \sigma w_t \label{sv1}\\
y_t &= \beta \exp\bigl\{\half x_t\bigr\}\epsilon_t, \label{sv2}
\end{align}
where $x_0 \sim \N(\mu, \frac{\sigma^2}{1-\phi^2})$, $w_t \distiid \N(0, 1)$, 
and $\epsilon_t \distiid \N(0,1)$ are all independent processes.
The volatility process $x_t$ is not observed directly, but only through the observations, $y_t$. 
 The constant $\mu$ is sometimes called a leverage effect and hence the model is also called the \emph{SV model with leverage} when $\mu\ne 0$.   The detailed econometric properties
  of the model are discussed in  \citet{Shephard1996} and  \citet{Taylor1994, Taylor2007}.

The model \eqref{sv1}--\eqref{sv2} is a nonlinear state space model, and Bayesian analysis
of such models can be approached via particle Gibbs methods; e.g., \citet[][Chap.~12]{Douc2014}.
Early MCMC approaches to the problem may be found in \citet{carlin1992}, \citet{Kim1998},
\citet{Jacquier1994}, and \citet{Taylor1994}.

Let $\Theta = (\mu, \beta,  \phi, \sigma)$ represent the parameters, denote
the observations as $\chunk{y}{1}{n}=\{y_1,\dots,y_\T\}$, and the states (log-volatility) by 
$\chunk{x}{0}{n} =\{x_0, x_1,\dots, x_\T\}$,
with $x_0$ being the initial state.
To  run a full Gibbs sampler, we alternate between
sampling model parameters and latent state sequences from their respective full conditional distributions.
Letting $\p(\cdot)$ denote a generic density, we have the following:
\begin{procedure}[Generic Gibbs Sampler for State Space Models]\label{proc:full-gibbs} 
 \begin{enumerate}[(i)]  \setlength{\itemsep}{0pt}
 \item\label{gibbsnum1bayessection}  Draw $\Theta^\prime \sim \p(\Theta \mid \chunk{x}{0}{n}, \chunk{y}{1}{n})$
 \item \label{gibbsnum2bayessection}  Draw $\chunk{x}{0}{n}^\prime \sim \p(\chunk{x}{0}{n} \mid \Theta^\prime, \chunk{y}{1}{n})$
 \end{enumerate}
\end{procedure}
\autoref{proc:full-gibbs}-\eqref{gibbsnum1bayessection}
is  generally 
much easier because it conditions on the complete data $\{\chunk{x}{0}{n}, \chunk{y}{1}{n}\}$.
\autoref{proc:full-gibbs}-\eqref{gibbsnum2bayessection} amounts to sampling from the joint smoothing distribution of the latent state sequence and is  generally difficult. 
  For linear Gaussian models, however, both parts of \autoref{proc:full-gibbs} are
 relatively easy to perform \citep[][Chap.~6]{fruhwirth1994, carter1994, Shumway2017}.

For nonlinear models, \autoref{proc:full-gibbs}-\eqref{gibbsnum2bayessection} can be
performed using particle methods. 
 \citet{DelMoral1996} introduced  the particle filter to sample the hidden states together from the conditional distribution. However,  particle filtering suffered from the path degeneracy, which
 makes sampling unreliable for long   time series   as  mentioned in \citet{Doucet2000}.   
 To avoid  path degeneracy, several resampling methods were introduced in late 1990s
 \citep[e.g.,][]{Doucet2000, liu1998}. 
  While the Forward Filter Backward Simulator (FFBSi) and the Forward Filter Backward Smoother (FFBSm) were introduced in  \citet{Doucet2000} and \citet{Godsill2004} to handle path degeneracy, the techniques 
  required the generation of many particles and resulted in an  approximation to the desired posterior distribution
  rather than yielding draws from the posterior distribution of interest.  

 \citet{Andrieu2010} introduced the particle Markov chain Monte Carlo (PMCMC) method,
 which proposed a conditional particle filter (CPF) to ease the difficulty of
 performing \autoref{proc:full-gibbs}-\eqref{gibbsnum2bayessection}.  The CPF
 is invariant in the sense that  the kernel leaves  $\p(\chunk{x}{0}{n} \mid \Theta^\prime, \chunk{y}{1}{n})$
 invariant; that is, all elements of the chain have the target distribution.
  CPF, however, suffers from the path degeneracy 
   and   works well only for short time series; otherwise, it is necessary to generate an 
   extremely large number    of particles.
   One way to solve this problem  involved a backward simulation sweep \citep{whiteley2010}. 
   However, the method is also computationally expensive.
   On the other hand, \citet{Lindsten2014} introduced a CPF with ancestral sampling (CPF-AS). The 
   addition of ancestral sampling
   improved on the problem path degeneracy while being robust to the number of particles generated.
  In fact, the method works very well with a small number of particles  and consequently
  is an invariant and efficient particle filter.

   The benefits of using CPF-AS to overcome the
  difficulty  of performing \autoref{proc:full-gibbs}-\eqref{gibbsnum2bayessection} is demonstrated
   using a number of examples    in \citet{Lindsten2014} and in
   \citet[][Chap.~12]{Douc2014}, where   \autoref{proc:full-gibbs}
  is called   Particle Gibbs with  Ancestor Sampling (PGAS) when CPF-AS is used for step 
  \autoref{proc:full-gibbs}-\eqref{gibbsnum2bayessection}.

As previously stated, step \autoref{proc:full-gibbs}-\eqref{gibbsnum1bayessection} is typically 
the easier step.  Usually, one puts normal  priors
   on  the leverage and (or beta priors) on the regression parameter(s), and inverse gamma priors on the scale parameters.
That is, current methods proceed as if $\phi$ is a regression parameter and $\sigma$ is a scale parameter
and this treatment  is what leads to the inefficiency for this particular model.
   The problem for SV models is that $\phi$  behaves   like a scale parameter as well as  a regression parameter. 
For example, the autocorrelation function (ACF) of
$\{ y_{t}^{2}\}$  is given by
\begin{equation}
\label{eq:acfSV}
\PCor(y^2_t,y^2_{t+h})=\frac{\exp(\sigma_{x}^{2}\phi^{h})-1}{\kappa_\epsilon\exp(\sigma_{x}^{2})-1} \,,
\quad h = 1,2, \dots\,,
\end{equation}
where $\kappa_\epsilon$ is the kurtosis of the noise, $\e_t$
and $\sigma_{x}^2 = \sigma^2/(1-\phi^2)$.
For SV models,   the ACF values are small  
 and the decay rate as a function of lag is less than exponential and somewhat linear. This means that
if you specify  values for $\phi$  but allow us to control $\sigma$ (and consequently $\sigma_x$),
we can make the model ACF to look approximately the same no matter which values of
$\phi$ are chosen.  This is accomplished by moving
$\phi$ and $\sigma$ in opposite directions.
Another way of looking at the problem is to let (with $\mu=0$ and $\beta=1$)
 $\xi_t = \frac{1}{2\sigma_x} x_t$ and $\zeta_t = \half w_t$ so we may write 
  \eqref{sv2} as
\begin{equation}\label{sv2a}
y_t = \E^{\sigma_x \phi  \xi_{t-1}} \E^{\sigma \cdot  \zeta_t} \epsilon_t \,,
\end{equation}
noting that $\xi_{t-1}$ and $\zeta_t$ are independent stationary $\half \N(0,1)$s.
It is clear from \eqref{sv2a} that $\sigma$ and $\phi$ are scale parameters of the 
$\xi_t$ process and $\sigma$ is a scale parameter of the $\zeta_t$ noise process;
we see  that we can keep the scale of the data approximately the same by
moving $\phi$ and $\sigma$ in opposite directions.

\begin{figure}[!tb]
\centerline{\includegraphics*[scale=.6]{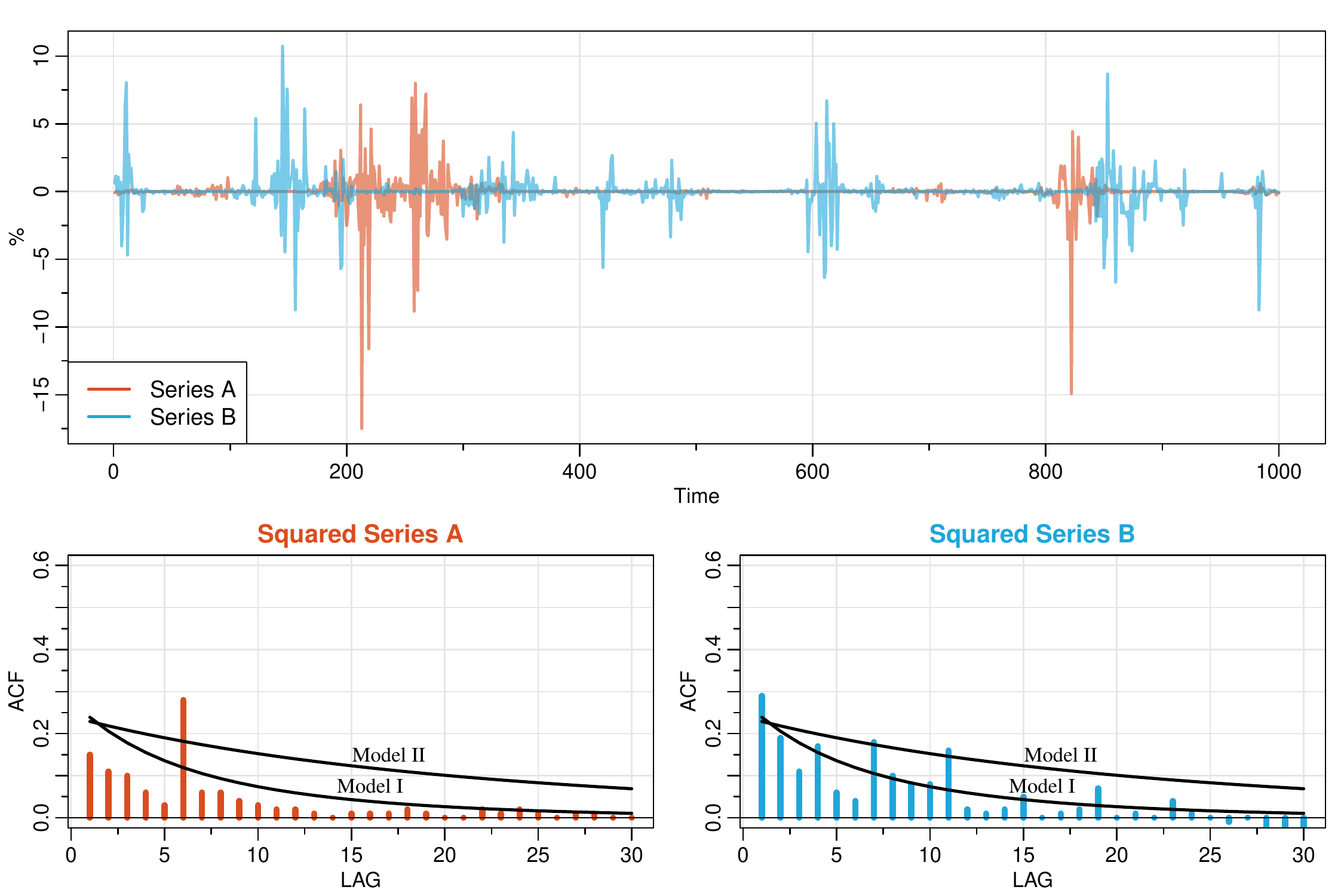}}
\caption{{\sc top:} Two data sequences (A and B) of length 1000 generated from  different 
two-parameter SV models, 
I: $\phi = .92$, $\sigma=1.5$ and II:  $\phi = .97$, $\sigma=1$.
{\sc bottom:} The ACF of each generated  series squared (A and B) and the theoretical
ACFs of SV  models I and II as lines.
The model (I or II) generating each series (A or B) is identified in the text.
\label{fig:compare} }
\end{figure}

For example, 
\autoref{fig:compare} shows two data sequences (A and B) of length 1000 generated from two different SV models,
\eqref{sv1}--\eqref{sv2}, with ($\mu=0$, $\beta=.1$)  
{\sc Model I:} $\phi = .92$, $\sigma=1.5$, and {\sc Model II:} $\phi = .97$, $\sigma=1$.
The ACF of each generated  series squared (A and B) and the theoretical
ACFs of SV models I and II as lines. 
While the AR parameter, $\phi$, is very different in each model, the simulated series look
very much the same.  In fact, counterintuitively,
series A corresponds to {\sc Model II:} $\phi = .97$, $\sigma=1$, and
series B corresponds to {\sc Model I:}  $\phi = .92$, $\sigma=1.5$.

While  CPF-AS can improve the mixing of the sampler for SV models, there are problems with
 slow convergence as noted by several authors \citep[e.g.,][]{Chib1996, Kim1998}
 because of the relationship between the parameters as previously noted.
 The problem persists to this day as one can see in  the
 vignettes of the \R\ package {\tt stochvol} \citep{kastner2019}.
 The inefficiency of the sampler is due to the fact that (in the two-parameter model)
 \autoref{proc:full-gibbs}-\eqref{gibbsnum1bayessection} is typically carried out
 in two steps  by drawing from the univariate posteriors
 $\p(\phi\mid \sigma, \chunk{x}{0}{n}, \chunk{y}{1}{n})$ and
 $\p(\sigma\mid \phi, \chunk{x}{0}{n}, \chunk{y}{1}{n})$.

As an example, we performed particle Gibbs with ancestral sampling (PGAS) on Series B ($n=1000$) shown in \autoref{fig:compare}, which was
generated from Model I,
\[x_t = .92 x_{t-1} + 1.5 w_t \quad  \text{and} \quad
y_t = .1\exp\bigl\{\half x_t\bigr\}\epsilon_t\,. \]
For \autoref{proc:full-gibbs}-\eqref{gibbsnum1bayessection}
we used  normal and inverse gamma priors for $\phi$ and $\sigma^2$, respectively.
That is, with prior $\sigma^2 \sim {\rm IG}(a_0/2, b_0/2)$,
where IG denotes the inverse (reciprocal) gamma distribution,
\begin{equation}\label{eq:inv_gamma}
\sigma^2 \mid \phi, x_{0:n},  y_{1:n}\sim
  {\rm IG}\Bigl(\tfrac{1}{2} (a_0+n+1),\ \tfrac{1}{2} \bigl\{ {b_0} +
         \sum_{t=1}^n [x_t - \phi x_{t-1}]^2  \bigr\}\Bigr)\,. 
\end{equation}         
With prior  $\phi \sim\N(\mu_\phi, \sigma_\phi^2)$, we have
$(\phi \mid \sigma,  x_{0:n},  y_{1:n} )\sim \N(Bb,B)$, where
\begin{equation}
B^{-1} = \frac{1}{\sigma^2_\phi} +  \frac{1}{\sigma^2}
 \sum_{t=1}^n {x_{t-1}^2}\,, \qquad
b = \frac{\mu_\phi}{\sigma^2_\phi} +  \frac{1}{\sigma^2}
  \sum_{t=1}^n  x_t x_{t-1}\,.
\end{equation} 
For the sake of exposition, we held $\mu$ and $\beta$ fixed at their given values
of $0$ and $.1$, respectively.
In addition, the values for $\sigma$ are larger than is typical for actual data,
but the large values help emphasize the problem.

For \autoref{proc:full-gibbs}-\eqref{gibbsnum2bayessection}, we used CPF-AS (\autoref{proc:cpfas})
with $N=20$ particles; details are given in the 
next section.
This example is similar to the experiment discussed in \citet[][Sec.~7.1]{Lindsten2013}, however  we
use simulated data so that we know the model is correct (and does not add to convergence problem considerations).

\begin{figure}[!tb]
\centerline{\includegraphics*[scale=.6]{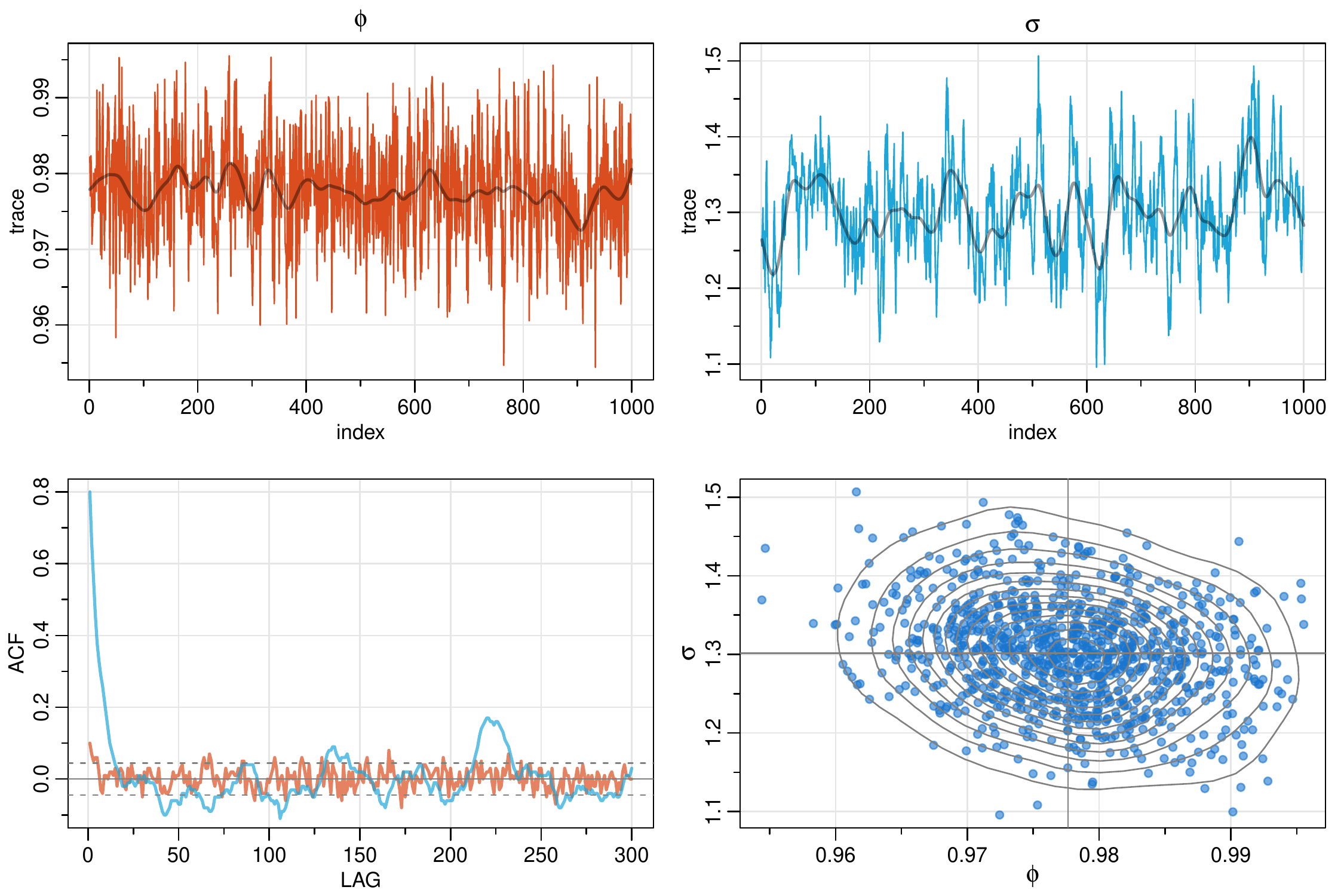}}
\caption{Particle Gibbs with individual parameter sampling
on Series B shown in \autoref{fig:compare}.
The actual parameters are $\phi=.92$ and $\sigma=1.5$.
{\sc top:}  Separate draws of $\phi$ and $\sigma$ in order
along with a lowess smooth to emphasize the cyclic behavior 
of the draws. 
{\sc bottom:} The sample ACF of the traces (left),
 and  a scatterplot (right) of the pairs of values in each draw
 with the posterior means of $.978$ and  $1.303$ highlighted.
\label{fig:simulate_pgas} }
\end{figure}

\autoref{fig:simulate_pgas} shows the results of the experiment.  The top row shows the
draws of $\phi$ and $\sigma$ in order, the bottom left shows the sample ACF of the traces,
 and on the right there is a scatterplot of the pairs of values in each draw.
One sees the slow convergence problem reported in \citet{Chib1996} and \citet{Kim1998}.
 The sampling procedure with CPF-AS  ameliorates the slow convergence problem to some degree,
 but is not a remedy because of relationship between the parameters seen in the scatterplot
 and previously discussed leads to a type of meandering through the sample spaces of the parameters.
 In fact, the sample paths look cyclic as emphasized by smoothing the traces.


To improve the efficiency of the algorithm, we propose a new method for  SV models by employing a bivariate  prior and sampling the parameters jointly from
 $\p(\phi, \sigma\mid  \chunk{x}{0}{n}, \chunk{y}{1}{n})$.
 A random walk Metropolis   algorithm is used to implement the new method. The new method reduces the parameter inefficiencies significantly. We also introduce an adaptive MCMC  to overcome the difficulty of choosing  the proposal distribution, if necessary. In addition, we extend our new method to the multivariate stochastic volatility (MSV) model.

\section{Particle Filtering}\label{sec:partfilter}

In this section, we review the particle method  used to perform step 
 \autoref{proc:full-gibbs}-\eqref{gibbsnum2bayessection}. The goal is to
 repeatedly draw
 an entire  state sequence from the posterior
  $\p(\chunk{x}{0}{n} \mid \Theta, \chunk{y}{1}{n})$.  To ease the notation,
  we will drop the conditioning arguments in this section. 
  Many of the details (along with references) for this section  may be found in \citet[][Part~III]{Douc2014}.

For notation, we will denote the proposal density by $\q(\cdot)$,
the target density by $\p(\cdot)$, and the importance function 
(unnormalized weight)
by $\om =\p/\q$.  Particle filtering is based on sequential
importance sampling and uses the fact that the states are Markov.
 
\begin{procedure}[Particle Filter]\label{proc:particlefilter} 
 \begin{enumerate}[(i)]  
   \item  Initialize,   $t=0$: 
     \begin{enumerate} \setlength{\itemindent}{-1em} 
     \item Draw $\epart[0]{j} \sim \q(\cdot)$    for $j = \range{1}{N}$ 
     \item Compute weights $ \weightfunc[0](\epart[0]{j}) = \p(\epart[0]{j})/\q(\epart[0]{j})$ for $j = \range{1}{N}$
     \item Normalize the weights $ \weightfunc[0]^j  =  \weightfunc[0](\epart[0]{j}) \bigm/
     \sum_{i=1}^{N} \weightfunc[0](\epart[0]{i})$  
     \end{enumerate}
   \item  for  $t=\range{1}{n}$:  
     \begin{enumerate} \setlength{\itemindent}{-1em} 
     \item Draw $I_t^j \sim$ Discrete$(\{\om_{t-1}^i\}_{i=1}^N)$ for   $j = \range{1}{N}$
      \item Draw $\epart[t]{j} \sim \q(x_t\| x_{0:t-1}^{I_t^j} )$  for   $j = \range{1}{N}$ 
     \item Set $\epart[0:t]{j} = (\epart[0:t-1]{I_t^j}, \epart[t]{j})$ 
  and  $\ewght[t]{j} \propto \weightfunc[t](\epart[t-1]{I_t^j}, \epart[t]{j})$,
     for $j = \range{1}{N}$
     \end{enumerate}
   \end{enumerate} 
\end{procedure}

Every density shown is conditional on parameters $\Theta$ and data $\chunk{y}{1}{t}$ up to time $t$.
At the end of the procedure, we will have a sample of size $N$ from the 
target of interest, $\p(\chunk{x}{0}{n} \mid \Theta, \chunk{y}{1}{n})$.
\autoref{proc:particlefilter} fails without some additional considerations such as
a resampling step that was   described in \citet{gordon:salmond:smith:1993}
and subsequently improved by others,  and an auxiliary adjustment step as described in  \citet{pitt:shephard:1999}.
To keep the exposition simple, we will henceforth assume that resampling and/or auxiliary methods are
applied appropriately in the procedures rather than explicitly showing these
necessary steps.  Simple multinomial resampling is used in all our examples.

Particle filtering was improved by \citet{Andrieu2010}, who proposed the
conditional particle filter (CPF) as follows.

\begin{procedure}[Conditional Particle Filter $\lbrack$CPF$\rbrack$]\label{proc:cpf} 
  {\sc Input:} A sequence of conditioned particles $x_{0:n}^{\prime}$ to  fix a reference trajectory. 
  \begin{enumerate}[(i)] 
  \item Initialize,   $t=0$:
    \begin{enumerate}\setlength{\itemindent}{-1em} 
    \item Draw $\epart[0]{j} \sim \q(\cdot)$    for $j = \range{1}{N-1}$ (sample only $N-1$ particles) 
    \item  Set the $N$th particle, $x_0^N =   x'_0$.   
    \item Compute weights $\ewght[0]{j} \propto \weightfunc[0](\epart[0]{j}) $ for $j = \range{1}{N}$
    \end{enumerate}
  \item for  $t=\range{1}{n}$:  
    \begin{enumerate}\setlength{\itemindent}{-1em}  
    \item Draw $I_t^j \sim$ Discrete$(\{\om_{t-1}^i\}_{i=1}^N)$ for   $j = \range{1}{N-1}$
     \item Draw $\epart[t]{j} \sim \q(x_t\| x_{0:t-1}^{I_t^j})$  for   $j = \range{1}{N-1}$ 
     \item  Set $x_t^N =    x'_t$   
    \item  Set $\epart[0:t]{j} = (\epart[0:t-1]{I_t^j}, \epart[t]{j})$ 
 and  $\ewght[t]{j} \propto \weightfunc[t](\epart[t-1]{I_t^j}, \epart[t]{j})$,
    for $j = \range{1}{N }$  
    \end{enumerate}
  \end{enumerate}
\end{procedure}

As previously mentioned,   CPF is invariant but  suffers from  path
degeneracy.  For large $n$,  the sample sequences  will typically degenerate
to the conditional path except  for the end of the sequence.   This problem prevents proper
mixing, and a remedy was considered in
 \citet{Lindsten2014} where they suggested that
 the conditional sequence be randomized.  This led to the  
  CPF with ancestral sampling (CPF-AS) approach.

\begin{procedure}[Conditional Particle Filter with Ancestral Sampling -- $\lbrack$CPF-AS$\rbrack$]\label{proc:cpfas} 
   {\sc Input:} A sequence of conditioned particles $x_{0:n}^{\prime}$ as a reference trajectory. 
   \begin{enumerate}[(i)]
  \item Initialize,   $t=0$:
    \begin{enumerate}\setlength{\itemindent}{-1em}
    \item Draw $\epart[0]{j} \sim \q(\cdot)$    for $j = \range{1}{N-1}$ (sample only $N-1$ particles) 
    \item  Set the $N$th particle, $x_0^N =   x'_0$.   
    \item Compute weights $\ewght[0]{j} \propto \weightfunc[0](\epart[0]{j}) $ for $j = \range{1}{N}$
    \end{enumerate}
  \item for  $t=\range{1}{n}$: 
    \begin{enumerate}\setlength{\itemindent}{-1em} 
    \item Draw $I_t^j \sim$ Discrete$(\{\om_{t-1}^i\}_{i=1}^N)$ for   $j = \range{1}{N-1}$
     \item Draw $\epart[t]{j} \sim \q(x_t\| x_{0:t-1}^{I_t^j})$  for   $j = \range{1}{N-1}$ 
     \item  Set $x_t^N =    x'_t$   
    \item  Draw $I_t^{N}\sim$ Discrete$(\{\om_{t-1}^i\}_{i=1}^N)$~~ (ancestor sample) 
    \item  Set $\epart[0:t]{j} = (\epart[0:t-1]{I_t^j}, \epart[t]{j})$ 
 and  $\ewght[t]{j} \propto \weightfunc[t](\epart[t-1]{I_t^j}, \epart[t]{j})$,
    for $j = \range{1}{N }$ 
    \end{enumerate}
  \end{enumerate}
\end{procedure} 

The difference between CPF and CFP-AS is that the reference trajectory,
 $x_{0:n}^{\prime}$, is randomized in the ancestral sampling step.  
 This step improves the mixing of the sampler and is  robust to
   small number of particles ($N= 5$ -- $20$).  Moreover,
   the (randomization) ancestral sampling  step does not affect the invariance properties 
   of the sampler (\citealp[e.g., see][]{Lindsten2013, Douc2014}).


\section{Proposed Method for Univariate Models}\label{sec:proposed1}

In the SV model, \eqref{sv1}--\eqref{sv2},  $\beta$ and $\mu$ are not both needed.
In choosing which parameter to keep,   
\citet{Kim1998}  argued that allowing  $\mu$ to vary and fixing $\beta= \exp(\mu/2)$
 has a better interpretation from an economic point-of-view.
Henceforth, we follow their restriction on $\beta$ and allow $\mu$ to vary.


In \autoref{sec:problem}, we discussed the problems of applying particle   methods to SV models.
Although  CPF-AS  solves  some  of the slow convergence problems reported by  \citet{Kim1998}, we still observe slow convergence caused by the high negative correlation between $\phi$ and $\sigma^2$. In this section, we suggest a new method to improve the convergence. 

The intuition of our method can be seen in the simulation displayed in \autoref{fig:simulate_pgas}.
That is, rather than sample $\phi$ and $\sigma$ individually, it would be better to sample
them at the same time.  That is, in the generic  Gibbs sampler, we accomplish
\autoref{proc:full-gibbs}-\eqref{gibbsnum1bayessection}
by sampling directly from
$\p(\phi, \sigma \mid \chunk{x}{0}{n}, \chunk{y}{1}{n})$ 
rather than sampling each parameter separately. As will be seen,
this change brings a big improvement to the mixing and convergence problems for SV models.

To accomplish this goal, we put a  bivariate normal prior with a negative correlation coefficient
on the pair $\Theta=(\phi, \sigma)$,
\begin{equation}\label{bi_normal}
    \left(\begin{array}{c} \phi \\ \sigma  \end{array}\right) \sim \N_2 \Biggl( \biggl[\begin{array}{c} \mu_{\phi} \\ \mu_{q}  \end{array} \biggr], \begin{bmatrix} \sigma_{\phi}^2 & \rho\sigma_{\phi}\sigma_{q} \\ \rho\sigma_{\phi}\sigma_{q} &  \sigma_{q}^2 \end{bmatrix} \Biggr),
\end{equation}
where typically, $\rho<0$. 
Allowing possible negative values for $\sigma$ is an old trick used in optimization to avoid constraints on the 
parameter space. The trick is reasonable because $\sigma^2$ will always be non-negative and has
 a  scaled chi-squared prior distribution. 
In addition,
as is seen in \autoref{fig:simulate_pgas}, a bivariate normal prior is sensible.
Note that we have changed the notation slightly by excluding $\mu$ from
$\Theta=(\phi, \sigma)$ because it may be sampled separately if necessary.
 
To accomplish \autoref{proc:full-gibbs}-\eqref{gibbsnum1bayessection},
note that,
\be
\dens(\param \mid  \chunk{x}{0}{n}, \chunk{y}{1}{n})  \propto  \pi(\param)\, \p(x_0\mid \param)\,
 \prod_{t=1}^n  \p(x_t \mid x_{t-1}, \param) \, \dens(y_t \mid x_{t}, \param)\,,
\label{5.9.8.2}
\ee
where $\pi(\param)$ is the prior on the parameters.
For the generic state space model, the parameters are often taken to be
conditionally independent with distributions from standard parametric families (at least
as long as the prior distribution is conjugate relative to the  model specification).
In this case, however, we must work
with non-conjugate models, and one option is to replace 
\autoref{proc:full-gibbs}-\eqref{gibbsnum1bayessection}
with a Metropolis-Hastings step, which is feasible because the complete data density
 $\dens(\param, \chunk{x}{0}{n}, \chunk{y}{1}{n})$ can be evaluated pointwise.
  
Under these considerations, for the  SV model in \eqref{sv1}--\eqref{sv2}, we have 
\begin{align} 
\p(\Theta \mid \mu, x_{0:\T}, y_{1:\T}) &\propto \pi(\Theta) \p(x_0\mid \param)
 \prod_{t=1}^n  \p(x_t \mid x_{t-1}, \param) \nonumber \\
&\propto 
    \exp  \Biggl\{ 
    -\frac{1}{2(1-\rho^2)}\biggl[ \frac{(\phi-\mu_{\phi})^2}{\sigma_{\phi}^2}+\frac{(\sigma-\mu_{\sigma})^2}{\sigma_q^2}-\frac{2\rho(\phi-\mu_{\phi}) (\sigma-\mu_{\sigma})}{\sigma_{\phi}\sigma_{q}} \biggr] 
    \Biggr\} \nonumber \\
&\cdot \frac{\sqrt{1-\phi^2}}{\sigma}\exp\biggl\{-\frac{(x_0-\mu)^2}{2\sigma^2/(1-\phi^2)}\biggr\}\prod_{t=1}^{\T}\frac{1}{\sigma} \exp\biggl\{ -\frac{[(x_t-\mu)-\phi(x_{t-1}-\mu)]^2}{2\sigma^2}\biggr\}\nonumber \\
&\propto 
    \exp  \Biggl\{ 
    -\frac{(\phi-\mu_{\phi})^2\sigma_{q}^2+(\sigma-\mu_{\sigma})^2\sigma_{\phi}^2-2\rho\sigma_{\phi}\sigma_{q}(\phi-\mu_{\phi}) (\sigma-\mu_{\sigma})}{2(1-\rho^2)\sigma_{\phi}^2\sigma_{q}^2} 
    \Biggr\} \nonumber \\
&\cdot
    \frac{\sqrt{1-\phi^2}}{\sigma^n}\exp\biggl\{ - \frac{(1-\phi^2)(x_0-\mu)^2+\sum_{t=1}^\T[(x_t-\mu)-\phi(x_{t-1}-\mu)]^2}{2\sigma^2} \biggr\}.  \label{eq:acceptance_prob}
\end{align}

As previously suggested, we  use a random walk Metropolis step to sample
 $\Theta = (\phi,\sigma)$ simultaneously from the target posterior distribution $\p(\Theta \mid \mu, x_{0:n},y_{1:n})$
 given in \eqref{eq:acceptance_prob}. 
This approach involves choosing a tuning parameter 
to control the acceptance probability.
 However, sometimes a good proposal distribution is difficult
 to choose because both the size and the spatial orientation of the proposal distribution should be
 considered. 
We have found that, for large samples, the use of an adaptive method 
can help with the problem.
We suggest using a  technique that was presented in \citet[][Alg.~4]{Andrieu:2008}
and is displayed in \autoref{proc:rwmh} for our case.

\begin{procedure}[Adaptive Normal Random Walk Metropolis]\label{proc:rwmh}
    {\sc Input:} Initial value, $\Theta_0$, and an initial bivariate normal proposal 
    distribution  $\N_2(\mu_0, \lambda_0\Sigma_0)$. \\
\noindent On iteration $j+1$, for $j=0,1,2,\dots$,   
\begin{enumerate}[(i)]  \setcounter{enumi}{0}  
    \item \label{rwmh_step1} Draw  $\vartheta \sim \N_2(\Theta_j,\lambda_j \Sigma_j)$  
      and set $\Theta_{j+1} = \vartheta$ with probability 
       $\alpha_{j+1} = \frac{g(\vartheta)}{g(\Theta_j)}\wedge 1$, where 
           $ g(\Theta)$ is given on the {\sc rhs} of \eqref{eq:acceptance_prob}.
        Otherwise, set    $\Theta_{j+1}=\Theta_{j}$
    \item \label{rwmh_step2} [Optional]\  Update 
            \begin{align}
                \log(\lambda_{j+1}) &= \log(\lambda_j) +  \gamma_{j+1}[\alpha_{j+1} - \alpha_\star], \label{equ:AM_update1}\\
                \mu_{j+1} &= \mu_j + \gamma_{j+1}(\Theta_{j+1} - \mu_j), \label{equ:AM_update2}\\
                \Sigma_{j+1} &= \Sigma_j + \gamma_{j+1}[(\Theta_{j+1} - \mu_j)(\Theta_{j+1} - \mu_j)' - \Sigma_j], \label{equ:AM_update3}
            \end{align}
            where $\gamma_{j}$ is a scalar nonincreasing sequence of positive  step lengths such that $\sum_{j=1}^{\infty} \gamma_j = \infty$ and $\sum_{j=1}^{\infty} \gamma_j^{1+\delta} < \infty$ for some $\delta>0$;  $\alpha_\star$ is the expected acceptance rate for the algorithm.

\end{enumerate}
\end{procedure}

Optionally, one may fix $\lambda_j$ and $\Sigma_j$ and skip step \autoref{proc:rwmh}-\eqref{rwmh_step2}
if it is not necessary. 
The optional part makes the algorithm non-Markovian, however, it
 can adapt continuously to the target distribution. Both the size and the spatial orientation of the proposal distribution will be adjusted by the adaptation procedure. Also, \autoref{proc:rwmh} is straightforward to implement and to use in practice.
 There are no extra computational costs because only a simple recursion formula for the covariances involved. The 
  algorithm starts by using the accumulating information from the beginning of the sampling and it ensures that the search becomes more efficient at an early stage of the sampling.  \citet{haario2001} establish that the adaptive MCMC algorithms do indeed have the correct ergodicity properties.

If the leverage parameter, $\mu$, is included in the model,
 using a diffuse prior \citep[e.g., see][]{Kim1998}, we have
\begin{equation}\label{sample_mu}
    \mu \mid \Theta, x_{0:n}, y_{1:n} \sim N(\nu_\mu, \sigma_\mu^2)
\end{equation}
where
\begin{equation*}
    \nu_\mu = \sigma_\mu^2 \bigg\{ \frac{1-\phi^2}{\sigma^2}x_0 + \frac{1-\phi}{\sigma^2}\sum_{t=1}^{\T}(x_{t}-\phi x_{t-1})\bigg\}
\end{equation*}
and
\begin{equation*}
    \sigma_\mu^2 = \frac{\sigma^2}{n(1-\phi)^2+(1-\phi^2)}.
\end{equation*}
Recall that we are fixing $\beta=\exp(\mu/2)$.
Finally, our algorithm for the analysis of a univariate SV model is given in \autoref{proc:PGAS_SV1}.

\begin{algorithm}[t]
  \caption{Joint  Particle Gibbs for Univariate Stochastic Volatility Models  \label{proc:PGAS_SV1}}
   {\sc Input:}  Set the initial value of $\Theta^{[0]}=(\phi, \sigma)^{[0]}, \mu^{[0]},$ and $x_{0:n}^{[0]}$ arbitrarily.\\
    At iteration  $j=1,2,\dots $,
    \begin{enumerate}[(i)] 
    \item Draw $x_{0:\T}^{[j]}$ by CPF-AS, \autoref{proc:cpfas},
     conditioned on $x_{0:n}^{[j-1]}$ and $\Theta^{[j-1]}, \mu^{[j-1]}$.
   \item With $x_{0:n}^{[j]}$, generate $\Theta^{[j]} =(\phi, \sigma)^{[j]}$ 
    via \autoref{proc:rwmh} and draw $\mu^{[j]}$ from the posterior given in \eqref{sample_mu}
    under the current draws $x_{0:n}^{[j]}$ and $\Theta^{[j]}$. 
      \end{enumerate}
\end{algorithm}

\section{Examples \label{sec:examples_univariate}}

\subsection{Joint versus Individual Sampling for a Two Parameter Model}\label{subsec:examples1}

\begin{figure}[t]
\centering\includegraphics[scale=.6]{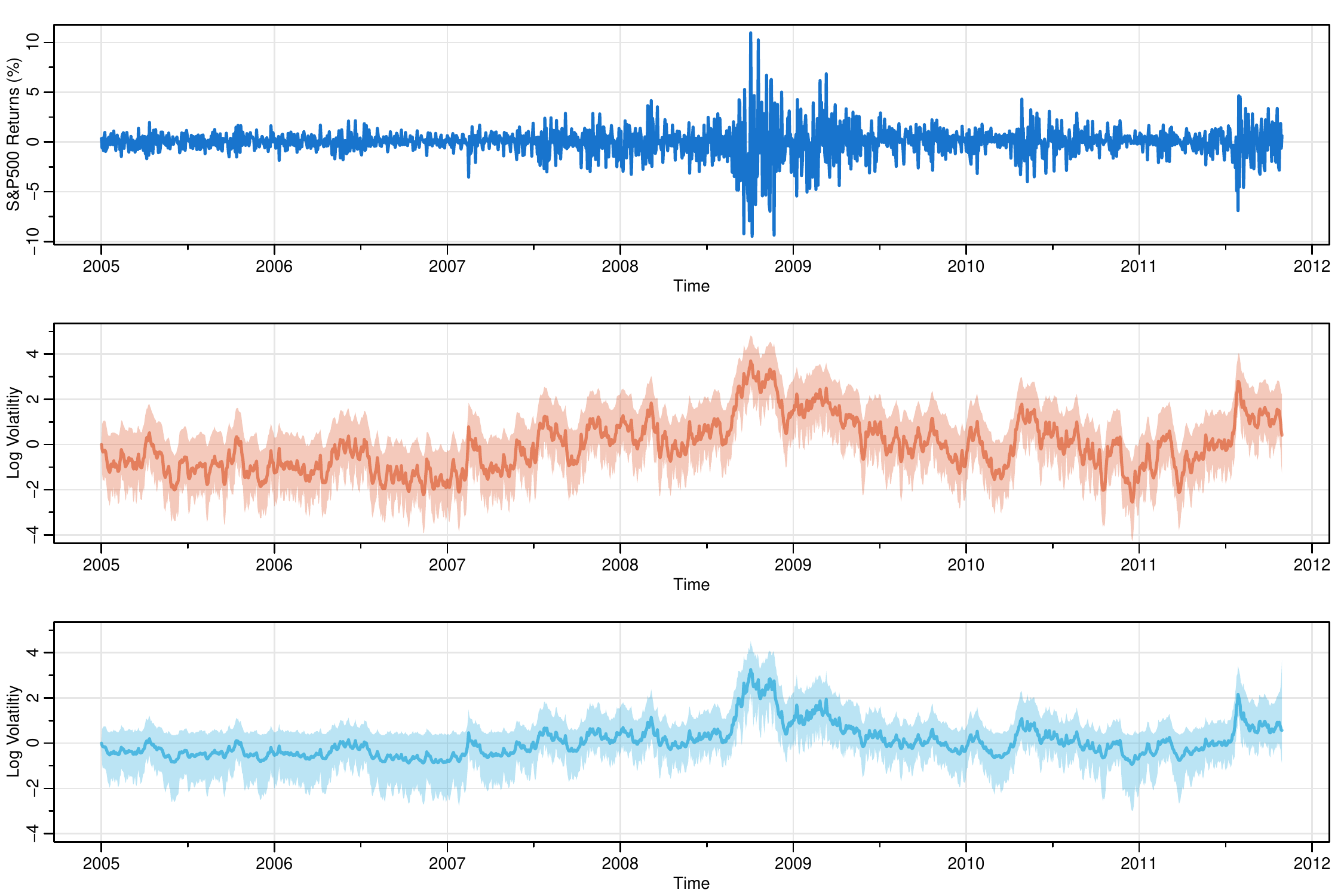}
\caption{{\sc top:}  Daily returns of the S\&P 500 from  January 2005 to October of 2011.
{\sc middle:} State process estimated posterior mean and pointwise 95\% credible intervals based
on the existing method of one-at-a-time parameter sampling.
{\sc bottom:} State process estimated posterior mean and pointwise 95\% credible intervals based
on the proposed method of joint parameter sampling, \autoref{proc:PGAS_SV1}.
The number of particles used in the particle filter  (\autoref{proc:cpfas})
for both methods was $N=20$.}
\label{fig:sp500_1}
\end{figure}

In this section, we fit a two-parameter model ($\mu=0$) to
  the daily returns of the S\&P 500 from  January 2005 to October of 2011  shown at
  the top of \autoref{fig:sp500_1}. The  data include the financial crisis of 2008.
We compare two particle Gibbs methods using CPF-AS (\autoref{proc:cpfas}) 
 to  sample the state process in both.
The standard (existing) method samples the parameters individually  
 by drawing from the univariate posteriors
 $\p(\phi\mid \sigma, \chunk{x}{0}{n}, \chunk{y}{1}{n})$ and
 $\p(\sigma\mid \phi, \chunk{x}{0}{n}, \chunk{y}{1}{n})$
 while our method  samples the
  parameters jointly  as described in 
 \autoref{proc:PGAS_SV1} holding $\mu$ at zero.

In each case, we used $N=20$ particles for the CPF-AS (\autoref{proc:cpfas})
and 5000 iterations after  a burnin of 100.
The posterior mean and a pointwise 95\% credible interval
of the draws of the state (log-volatility) process is shown
in \autoref{fig:sp500_1}. The middle plot shows the results for the existing method
and the bottom plot shows the results for our proposed method.
The results are similar, but the trace of the estimated process 
is smoother and less variable than existing method shown in the middle.

\begin{figure}[t]
\centering\includegraphics[scale=.6]{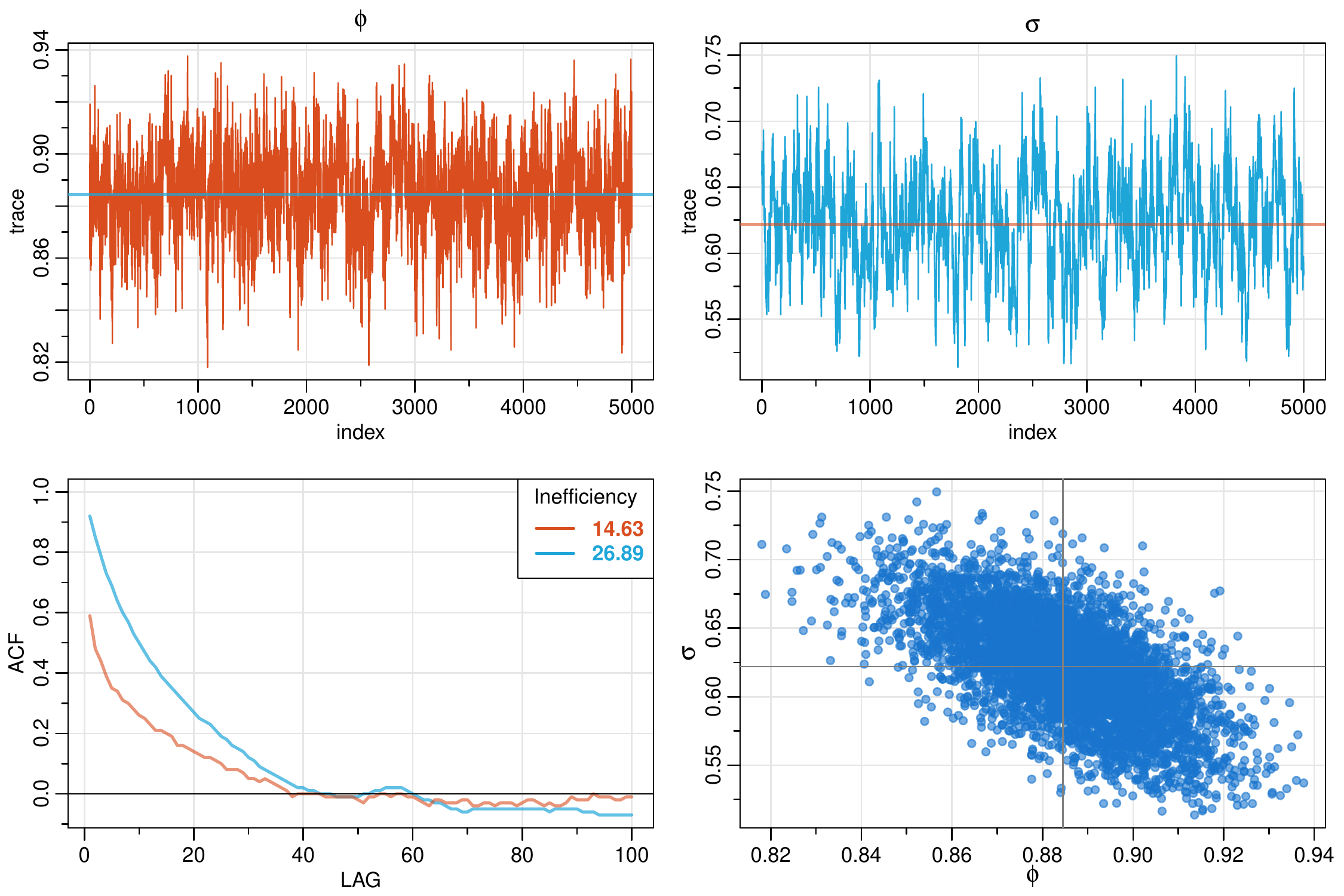}
\caption{{\bf (Individual Sampling)} S\&P 500,  results of the parameter estimation in a two-parameter model using the standard method 
of sampling $\phi$ and $\sigma$ separately. 
The top shows the traces of the 5000 sampled values after a burnin of 100. 
The number of particles used in the particle filter (\autoref{proc:cpfas})
was $N=20$.
The corresponding posterior means are $.88$ for $\phi$ and
$.62$ for $\sigma$.
The bottom left shows the sample ACFs of the traces 
and the estimated inefficiency measure as defined in \eqref{eq:IF}.
The bottom left shows a scatterplot of the sampled parameters, and exhibits
strong negative correlation.}
\label{fig:sp500_2}
\end{figure}

\autoref{fig:sp500_2} displays the results of the parameter estimation using the standard method 
of sampling $\phi$ and $\sigma$ separately. 
The top of the figure shows the traces of the sampled values after burnin. 
The corresponding posterior means are $.88$ for $\phi$ and
$.62$ for $\sigma$.
The bottom of the 
figure shows the sample ACFs of the traces and a scatterplot of the sampled values.  In addition,
the ACF plot displays the inefficiency measure as defined in \citet{geyer1992}.  The measure
was obtained using  Geyer's \R\ package, {\tt mcmc} \citep{geyer_mcmc_package}.  In particular,
to evaluate the mixing of sampler, we estimate  \textit{inefficiency}  defined as 
\begin{equation}\label{eq:IF}
    \text{IF} := 1 + 2 \sum_{i=1}^\infty \rho(i),
\end{equation}
where $\rho(i)$ is the autocorrelation function of the trace at lag $i$. 
The estimated inefficiencies are displayed with the sample ACFs of traces. 
We note again the slow convergence problem 
seen in the simulation example,  \autoref{fig:simulate_pgas}, and
reported in \citet{Chib1996} and \citet{Kim1998}.
Finally, the bottom right scatterplot shows the strong correlation
between the individually sampled parameters.

%
%

\begin{figure}[t]
\centering\includegraphics[scale=.6]{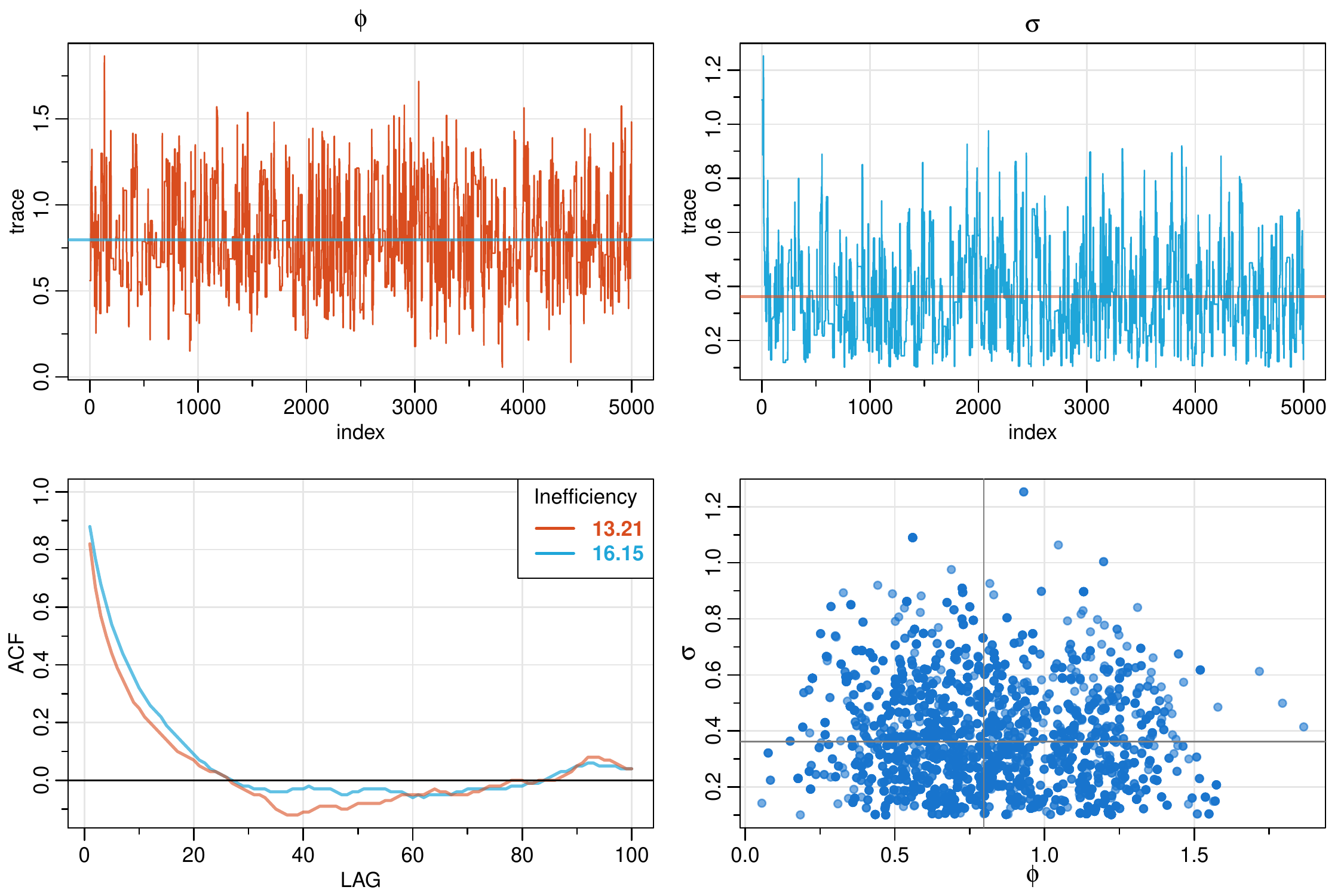}
\caption{{\bf (Joint Sampling)} S\&P 500, results of the parameter estimation in a two-parameter model using the proposed method
of sampling $\phi$ and $\sigma$ jointly, \autoref{proc:PGAS_SV1}. 
The top shows the traces of the 5000 sampled values after a burnin of 100. 
The number of particles used in the particle filter (\autoref{proc:cpfas})
was $N=20$.
The corresponding posterior means are $.80$ for $\phi$ and
$.36$ for $\sigma$.
The bottom shows the sample ACFs of the traces and a scatterplot of the sampled values,
which indicate the sample parameters are uncorrelated.  In addition,
 estimated inefficiency measures are improved over the counterparts in
 \autoref{fig:sp500_2}.}
\label{fig:sp500_3}
\end{figure}

\autoref{fig:sp500_3} displays the results of the parameter estimation using
our proposed method,  
\autoref{proc:PGAS_SV1},
sampling $\phi$ and $\sigma$ simultaneously.
The top of the figure shows the traces of the sampled values after burnin. 
The corresponding posterior means are $.80$ for $\phi$ and
$.36$ for $\sigma$.
The bottom of the 
figure shows the sample ACFs of the traces and a scatterplot of the sampled values,
which shows an improvement of the established method.
While the inefficiencies of estimating $\phi$ are similar,
the  inefficiencies of estimating $\sigma$ are much improved.
In addition, the scatterplot of the joint draws of the parameters 
shows the absence of correlation among the samples.

\subsection{Three Parameter Model}

Next, we fit a three parameter SV model to the S\&P 500 series using \autoref{proc:PGAS_SV1},
however,  the adaptive part of the Metropolis step,  
 \autoref{proc:rwmh}-\eqref{rwmh_step2}, was skipped.
 To keep the complexity low, we 
used only $N=10$ particles for sampling the states  (\autoref{proc:cpfas}),
 and then generated 2000 samples after a burnin of 100.
The acceptance rate was nearly optimal at $26.1\%$.
The entire estimation process took less than 4 minutes on a
workstation running Windows 10 Pro with 32GB of DDR3 RAM, an Intel i7-4770 CPU @ 3.40 GHz, and
using Microsoft R, version 3.5.2.

\begin{figure}[t]
\centering\includegraphics[scale=.65]{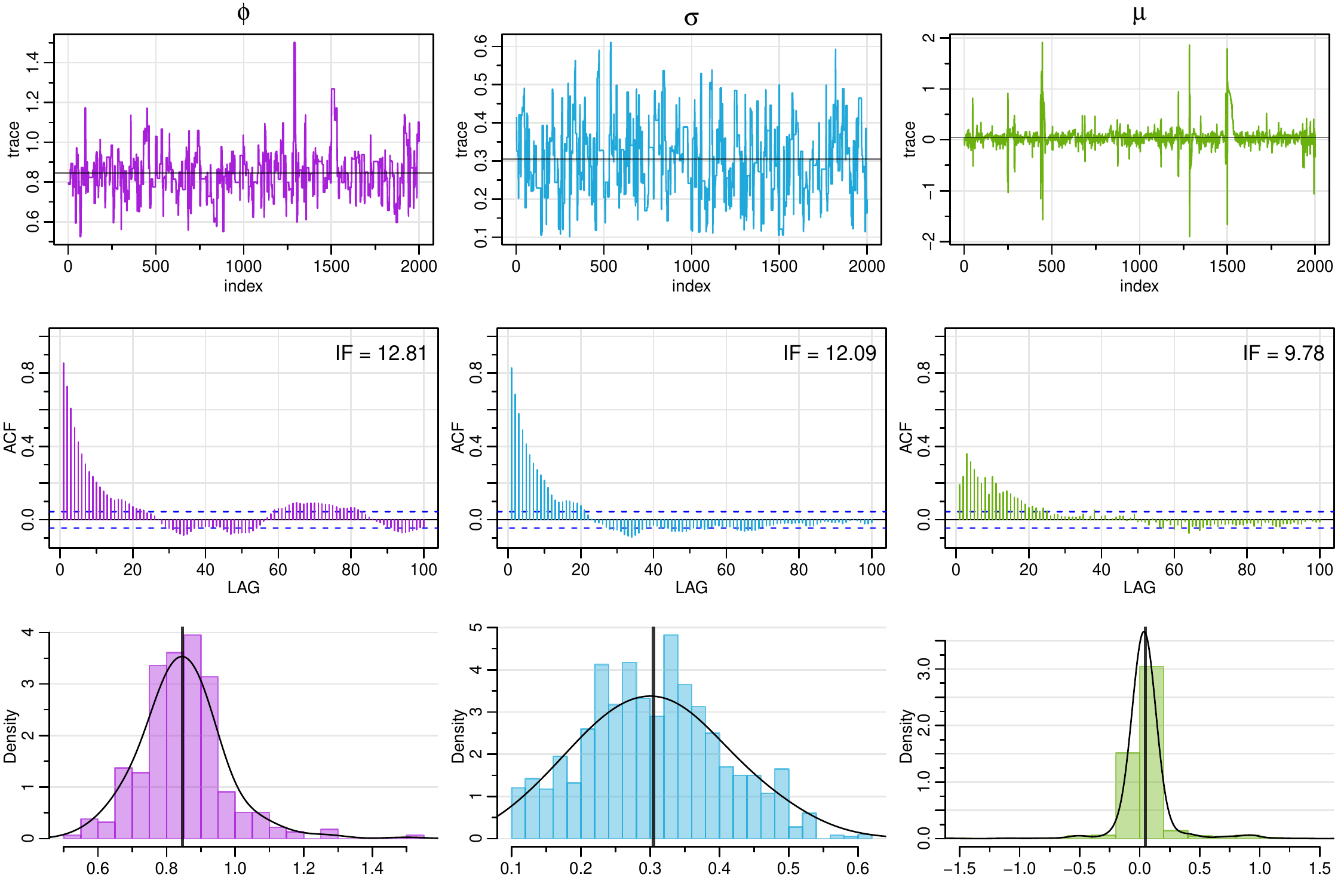}
\caption{S\&P 500, 
results of the parameter estimation in a three-parameter model using the proposed method
of sampling $\phi$ and $\sigma$ jointly, \autoref{proc:PGAS_SV1}. 
{\sc top:} Traces of the 2000 sampled values of the parameters after burnin.
{\sc middle:}  Sample ACF of the draws  along with the estimated inefficiency, \eqref{eq:IF}.
{\sc bottom:} Histogram of the results.  The posterior means are displayed in the figure
and were $.85$ for $\phi$, $.30$ for $\sigma$ and $.05$ for $\mu$.}
\label{fig:sp500_3parms}
\end{figure}

The results of the parameter estimation are shown in \autoref{fig:sp500_3parms}; the
results for the state estimation are similar to the lower plot of \autoref{fig:sp500_1}
and are not shown to save space.  The figure shows the trace of the draws (top row), the
sample ACF of the draws (middle row)  along with the estimated inefficiency, \eqref{eq:IF},
and a histogram of the results (bottom row).  The posterior means are displayed in the figure
and were $.85$ for $\phi$, $.30$ for $\sigma$ and $.05$ for $\mu$.
We note that the results are satisfactory even using this quick analysis. In addition,
it is apparent that the previous analysis based on the two-parameter model ($\mu=0$)
was reasonable.

%
%
%

\section{Multivariate Stochastic Volatility Model} \label{sec:mSV}
It is often reasonable to assume that similar assets are being driven by
the same volatility process.  In this case,  
 the multivariate stochastic volatility (MSV) model presented in \citet{Asai2006}
 can be used.  The model assumes a univariate volatility process is
 driving a number of similar assets and is given by, 
\begin{align}
x_t &= \phi x_{t-1}+  \sigma w_t \label{msv1_def1}\\
{y}_{it} &= \beta_i\exp\Bigl\{ \frac{x_t}{2} \Bigr\}\,\epsilon_{it}\,, 
\quad i=1,\dots, p , \label{msv_def2}
\end{align}
where the
$y_{it}$ are the returns of the $i$th asset, 
 $w_t \distiid  \N(0, 1)$, and
$\bm{\epsilon}_t = (\epsilon_{1t}, \dots,\epsilon_{pt})' \distiid \N_p(0, \Id)$. 
In this model, the leverage ($\mu$) is removed to avoid overparameterization  
and each $\beta_i$ is a scale parameter for the $i$th asset.

We can easily apply our proposed method to the MSV model.  That is,
as in the univariate case, 
we   put a bivariate normal prior on the state parameters, 
$\phi$ and $\sigma$. Then, because
   $\beta_i$, for $i=1,\dots, p$, is a scale parameter, a reasonable choice
is to use independent inverse gamma priors for $\beta_i^2$
as in \eqref{eq:inv_gamma}. That is, if
$\beta_i^2 \sim {\rm IG}(a_i/2, b_i/2)$,
then the posterior is
\begin{equation} \label{eq:beta_poseterior}
    \beta_i^2 \mid \Theta, x_{0:n},  y_{1:n}\sim 
      {\rm IG}\Bigl(\tfrac{1}{2} (a_i+n+1),\ \tfrac{1}{2} \Bigl\{ b_i +
             \sum_{t=1}^n \frac{y_{it}^2}{\exp(x_t)}\Bigr\}\Bigr)\,. 
\end{equation}         
Hence, in the MSV model, we can simply add a third step 
to \autoref{proc:PGAS_SV1}, which is to sample
$\beta_i^2$ from \eqref{eq:beta_poseterior} for $i=1,\dots, p$.
We summarize these steps in \autoref{proc:PGAS_MSV}.

\begin{algorithm}[t]
  \caption{Joint  Particle Gibbs for Multivariate Stochastic Volatility Models  \label{proc:PGAS_MSV}}
   {\sc Input:}  Set the initial value of $\Theta^{[0]}=(\phi, \sigma)^{[0]},$
   $\beta_{1:p}^{[0]}$,  and $x_{0:n}^{[0]}$ arbitrarily.\\
    At iteration  $j=1,2,\dots $,
    \begin{enumerate}[(i)] 
    \item Draw $x_{0:\T}^{[j]}$ by CPF-AS, \autoref{proc:cpfas},
     conditioned on $x_{0:n}^{[j-1]}$ and $\Theta^{[j-1]}$.
   \item With $x_{0:n}^{[j]}$, generate $\Theta^{[j]} =(\phi, \sigma)^{[j]}$ 
    via \autoref{proc:rwmh} and draw $\beta_{1:p}^{2\,[j]}$ from the posteriors given in \eqref{eq:beta_poseterior}
    under the current draws $x_{0:n}^{[j]}$ and $\Theta^{[j]}$. 
      \end{enumerate}
\end{algorithm}

\begin{figure}[t]
\centering\includegraphics[scale=.6]{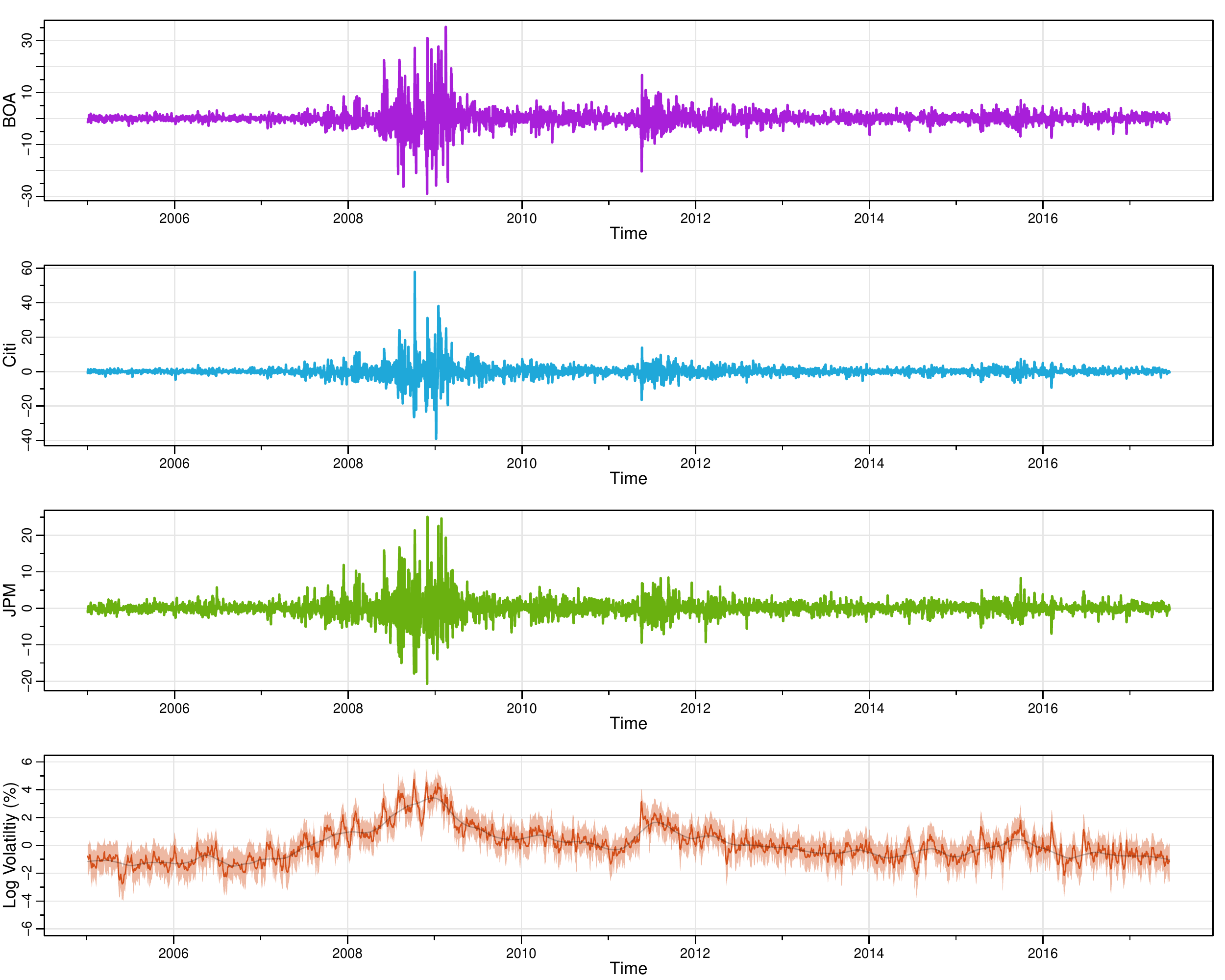}
\caption{The daily NYSE returns (as percentages) for three banks,
Bank of American (BOA), Citigroup (Citi), and 
J.P. Morgan (JPM) from January 2005 to November 2017.
{\sc bottom:} The  resulting posterior of the log-volatility 
based on \autoref{proc:PGAS_MSV}.
Shown are the posterior mean and a swatch displaying the
limits of $99\%$ of the sampled states.  We also display
a lowess smooth as a thin line to emphasize the volatility trend. }
\label{fig:MSVstate}
\end{figure}

For an example, we consider the daily NYSE returns for three banks,
Bank of American (BOA), Citigroup (Citi), and 
J.P. Morgan (JPM) from January 2005 to November 2017.
The data are displayed in \autoref{fig:MSVstate}; also shown is
the estimated log-volatility, which we describe shortly.

\begin{figure}[t]
\centering\includegraphics[scale=.6]{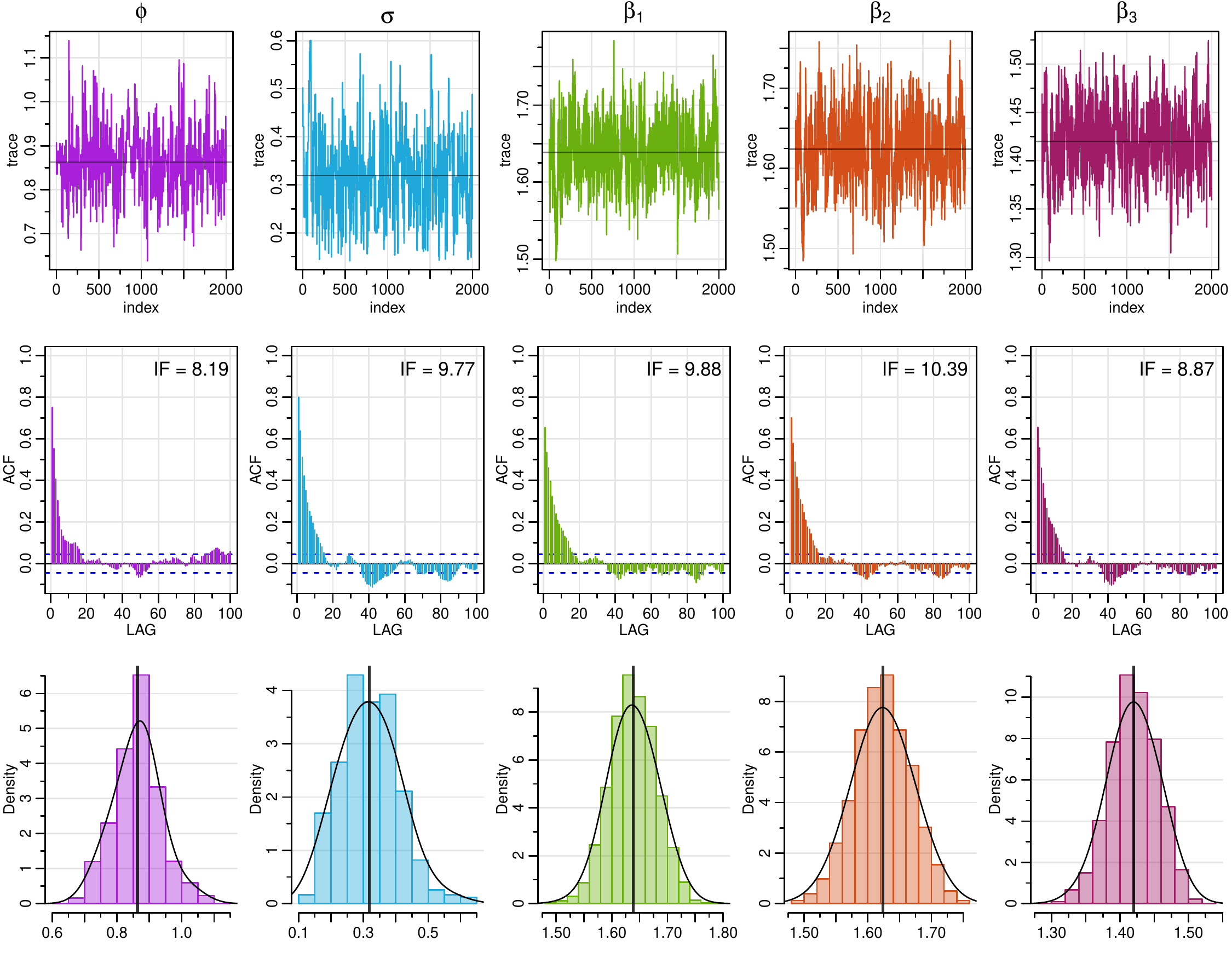}
\caption{For the bank returns data shown in \autoref{fig:MSVstate},
the results of fitting an MSV model based on \autoref{proc:PGAS_MSV}.
{\sc top:} The traces of the draws for each parameter and indicates the
posterior means, $.86$ for $\phi$, $.32$ for $\sigma$, and
$1.64$, $1.62$, and $1.42$ for the $\beta$s of
BOA, Citi, and JPM, respectively.
{\sc middle:} The sample ACFs of the traces
along with the inefficiencies.
{\sc bottom:} The estimated
posterior distributions of each parameter along with the
location of the posterior mean.}
\label{fig:MSVparms}
\end{figure}

We used \autoref{proc:PGAS_MSV} with $N=20$ particles
to generate 2000 draws after a burnin of 500 iterations.
The procedure was non-adaptive and the acceptance rate
was $29.8\%$.  The entire procedure took about 12 minutes
on the same machine mentioned in the other examples.
The parameter estimation summary is displayed in
\autoref{fig:MSVparms} and the display is similar to the
previous example.  The displays suggest that the algorithm
is mixing well.   The top shows the
traces of the draws for each parameter and indicates the
posterior means, $.86$ for $\phi$, $.32$ for $\sigma$, and
$1.64$, $1.62$, and $1.42$ for the $\beta$s of
BOA, Citi, and JPM, respectively.
The middle plot shows the sample ACFs of the traces
along with the inefficiencies.
The bottom row of \autoref{fig:MSVparms} displays the
posterior distributions of each parameter along with the
location of the posterior mean.

The resulting posterior of the log-volatility is shown
at the bottom of \autoref{fig:MSVstate}.  
Shown are the posterior mean and a swatch displaying 
pointwise  $99\%$ credible intervals.  We also display
a lowess fit as a thin line to emphasize the volatility trend.
Notice that the impending financial crisis of 2008 is visible
at least one year prior as the volatility starts a trend upwards
just prior to 2007. 
It seems that there is an advantage to using multiple similar sources
to estimate volatility.

%

\section{Conclusions}\label{sec:end}

The conditional particle filter with ancestral sampling (\autoref{proc:cpfas}) was
a breakthrough for analyzing nonlinear state space models by establishing a computationally efficient
method of sampling from the posterior of the hidden state trajectories, 
\autoref{proc:full-gibbs}-\eqref{gibbsnum2bayessection}.
The method works well for many cases if  
drawing from the posterior of the parameters, 
\autoref{proc:full-gibbs}-\eqref{gibbsnum1bayessection}, 
is not problematic. Unfortunately, this situation does not include the case
of stochastic volatility models because in the state equation,
the autoregressive parameter, $\phi$ and the noise variance, $\sigma^2$ in \eqref{sv1} have a tendency
to work in opposite directions.

Prior attempts to handle SV models had efficiency problems because 
$\phi$ was treated as a regression parameter while $\sigma$ was treated 
as a scale parameter. 
  Consequently, these parameters were sampled individually as they typically
  are in these situations.
For many state space models, this treatment of the problem is fine.  For
SV models however, this approach is an efficiency nightmare.

We have presented a method to overcome this problem by sampling the state
parameters jointly.  We used a bivariate normal distribution based on
the fact that it is easy to work with in that it captures the subtleties of
the relationship, but also, as seen in \autoref{fig:sp500_2}, $\phi$ and 
$\sigma$ live on ellipses.
While it is possible that a sampled pair yields values of  $|\phi|>1$ or $\sigma<0$,
it does not appear to be a problem.  For example, the state process is assumed to
be stationary, so realistically, one only needs $|\phi|\ne 1$, which will not happen
(with probability 1 in all but pathological cases). Also, sampled values of $\sigma^2$ will always be non-negative. 
  We do note that, even though $\sigma$ was small in our examples,
we never saw a negative value of $\sigma$.

Finally, we mention that we did not supply every particular numerical detail (e.g.,
hyperparameters and tuning parameters)
of our examples. Instead, for the sake of reproducibility, we supply the \R\ code for every example 
 on GitHub; see \cite{GitHub2019} for the {\sc url}.  
Additional information may also be found in \cite{Gongphd2019}.

\bibliographystyle{apalike}
\bibliography{reference}

\begin{thebibliography}{}

\bibitem[Andrieu et~al., 2010]{Andrieu2010}
Andrieu, C., Doucet, A., and Holenstein, R. (2010).
\newblock Particle {M}arkov chain {M}onte {C}arlo methods.
\newblock {\em Journal of the Royal Statistical Society: Series B (Statistical
  Methodology)}, 72(3):269--342.

\bibitem[Andrieu and Thoms, 2008]{Andrieu:2008}
Andrieu, C. and Thoms, J. (2008).
\newblock A tutorial on adaptive {MCMC}.
\newblock {\em Statistics and Computing}, 18(4):343--373.

\bibitem[Asai et~al., 2006]{Asai2006}
Asai, M., McAleer, M., and Yu, J. (2006).
\newblock Multivariate stochastic volatility: a review.
\newblock {\em Econometric Reviews}, 25(2-3):145--175.

\bibitem[Carlin et~al., 1992]{carlin1992}
Carlin, B.~P., Polson, N.~G., and Stoffer, D.~S. (1992).
\newblock A {M}onte {C}arlo approach to nonnormal and nonlinear state-space
  modeling.
\newblock {\em Journal of the American Statistical Association},
  87(418):493--500.

\bibitem[Carter and Kohn, 1994]{carter1994}
Carter, C.~K. and Kohn, R. (1994).
\newblock On {G}ibbs sampling for state space models.
\newblock {\em Biometrika}, 81(3):541--553.

\bibitem[Chib and Greenberg, 1996]{Chib1996}
Chib, S. and Greenberg, E. (1996).
\newblock Markov chain {M}onte {C}arlo simulation methods in econometrics.
\newblock {\em Econometric Theory}, 12(3):409--431.

\bibitem[Del~Moral, 1996]{DelMoral1996}
Del~Moral, P. (1996).
\newblock Non-linear filtering: Interacting particle resolution.
\newblock {\em Markov Processes and Related Fields}, 2(4):555--581.

\bibitem[Douc et~al., 2014]{Douc2014}
Douc, R., Moulines, E., and Stoffer, D.~S. (2014).
\newblock {\em Nonlinear Time Series: Theory, Methods and Applications with R
  Examples}.
\newblock CRC Press, Boca Raton.

\bibitem[Doucet et~al., 2000]{Doucet2000}
Doucet, A., Godsill, S., and Andrieu, C. (2000).
\newblock On sequential {M}onte {C}arlo sampling methods for {B}ayesian
  filtering.
\newblock {\em Statistics and Computing}, 10(3):197--208.

\bibitem[Fr{\"u}hwirth-Schnatter, 1994]{fruhwirth1994}
Fr{\"u}hwirth-Schnatter, S. (1994).
\newblock Data augmentation and dynamic linear models.
\newblock {\em Journal of Time Series Analysis}, 15(2):183--202.

\bibitem[Geyer, 1992]{geyer1992}
Geyer, C.~J. (1992).
\newblock Practical {M}arkov chain {M}onte {C}arlo.
\newblock {\em Statist. Sci.}, 7(4):473--483.

\bibitem[Geyer and Johnson, 2017]{geyer_mcmc_package}
Geyer, C.~J. and Johnson, L.~T. (2017).
\newblock {\em mcmc: Markov Chain Monte Carlo}.
\newblock R package version 0.9-5.

\bibitem[Godsill et~al., 2004]{Godsill2004}
Godsill, S.~J., Doucet, A., and West, M. (2004).
\newblock Monte {C}arlo smoothing for nonlinear time series.
\newblock {\em Journal of the American Statistical Association},
  99(465):156--168.

\bibitem[Gong, 2019]{Gongphd2019}
Gong, C. (2019).
\newblock {\em Particle Gibbs Methods in Stochastic Volatility Models}.
\newblock PhD thesis, University of Pittsburgh.

\bibitem[Gong and Stoffer, 2019]{GitHub2019}
Gong, C. and Stoffer, D.~S. (2019).
\newblock {Stochastic Volatility Models}.
\newblock \url{https://github.com/nickpoison/Stochastic-Volatility-Models/}.
\newblock [GitHub Repository].

\bibitem[Gordon et~al., 1993]{gordon:salmond:smith:1993}
Gordon, N., Salmond, D., and Smith, A.~F. (1993).
\newblock Novel approach to nonlinear/non-{G}aussian {B}ayesian state
  estimation.
\newblock {\em IEE Proc. F, Radar Signal Process.}, 140:107--113.

\bibitem[Haario et~al., 2001]{haario2001}
Haario, H., Saksman, E., and Tamminen, J. (2001).
\newblock An adaptive {M}etropolis algorithm.
\newblock {\em Bernoulli}, 7(2):223--242.

\bibitem[Jacquier et~al., 1994]{Jacquier1994}
Jacquier, E., Polson, N.~G., and Rossi, P.~E. (1994).
\newblock Bayesian analysis of stochastic volatility models.
\newblock {\em Journal of Business \& Economic Statistics}, 20(1):69--87.

\bibitem[Kastner and Hosszejni, 2019]{kastner2019}
Kastner, G. and Hosszejni, D. (2019).
\newblock {\em {stochvol}: Efficient Bayesian Inference for Stochastic
  Volatility ({SV}) Models}.
\newblock R package version 2.0.4.

\bibitem[Kim et~al., 1998]{Kim1998}
Kim, S., Shephard, N., and Chib, S. (1998).
\newblock Stochastic volatility: {L}ikelihood inference and comparison with
  {ARCH} models.
\newblock {\em The Review of Economic Studies}, 65(3):361--393.

\bibitem[Lindsten et~al., 2014]{Lindsten2014}
Lindsten, F., Douc, R., and Moulines, E. (2014).
\newblock Particle {G}ibbs with ancestor sampling.
\newblock {\em Journal of Machine Learning Research}, 15:2145--2184.

\bibitem[Lindsten et~al., 2013]{Lindsten2013}
Lindsten, F., Sch{\"o}n, T.~B., et~al. (2013).
\newblock Backward simulation methods for {M}onte {C}arlo statistical
  inference.
\newblock {\em Foundations and Trends in Machine Learning}, 6(1):1--143.

\bibitem[Liu and Chen, 1998]{liu1998}
Liu, J.~S. and Chen, R. (1998).
\newblock Sequential {M}onte {C}arlo methods for dynamic systems.
\newblock {\em Journal of the American Statistical Association},
  93(443):1032--1044.

\bibitem[Pitt and Shephard, 1999]{pitt:shephard:1999}
Pitt, M.~K. and Shephard, N. (1999).
\newblock Filtering via simulation: Auxiliary particle filters.
\newblock {\em Journal of the American Statistical Association},
  94(446):590--599.

\bibitem[Shephard, 1996]{Shephard1996}
Shephard, N. (1996).
\newblock Statistical aspects of {ARCH} and stochastic volatility.
\newblock {\em Monographs on Statistics and Applied Probability}, 65:1--68.

\bibitem[Shumway and Stoffer, 2017]{Shumway2017}
Shumway, R.~H. and Stoffer, D.~S. (2017).
\newblock {\em Time Series Analysis and Its Applications: With R Examples}.
\newblock Springer, New York, 4th edition.

\bibitem[Taylor, 1994]{Taylor1994}
Taylor, S.~J. (1994).
\newblock Modeling stochastic volatility: A review and comparative study.
\newblock {\em Mathematical Finance}, 4(2):183--204.

\bibitem[Taylor, 2008]{Taylor2007}
Taylor, S.~J. (2008).
\newblock {\em Modelling Financial Time Series}.
\newblock World Scientific, 2nd edition.

\bibitem[Whiteley et~al., 2010]{whiteley2010}
Whiteley, N., Andrieu, C., and Doucet, A. (2010).
\newblock Efficient {B}ayesian inference for switching state-space models using
  discrete particle {M}arkov chain {M}onte {C}arlo methods.
\newblock {\em arXiv preprint arXiv:1011.2437}.

\end{thebibliography}
\end{document}